\documentclass[acmsmall]{acmart}

\AtBeginDocument{%
  }

\usepackage{enumitem}
\usepackage{array}
\usepackage{wrapfig}
\usepackage{amsmath,amsfonts}
\usepackage[noend]{algpseudocode}
\usepackage{graphicx}
\usepackage{textcomp}
\usepackage{float}
\usepackage{listings}
\usepackage{xspace}
\usepackage{multirow}
\usepackage{amsthm}

\usepackage{balance}
\usepackage{algorithm}
\usepackage{algpseudocode}
\usepackage{colortbl}
\usepackage{subcaption}

\usepackage[skins]{tcolorbox}
\usepackage{xcolor}
\usepackage{multicol}
\usepackage{soul}
\usepackage{framed}
\usepackage{booktabs}
\usepackage{booktabs}
\usepackage{multirow}
\usepackage{siunitx}

\usepackage{algorithm}
\usepackage{tabularx}
\usepackage{booktabs}
\usepackage{color}
\usepackage{multirow}
\usepackage{listings}
\usepackage{pythonhighlight}
\usepackage{subcaption}
\usepackage{tikz}
\usetikzlibrary{positioning}
\usepackage{tabularx}
\usepackage{algpseudocode}
\usepackage{graphicx}
\usepackage{balance}
\usepackage{pgfplotstable}
\usepackage{tcolorbox}
\usepackage{lipsum}

\usepackage[normalem]{ulem}
\newcommand*\colourcheck[1]{%
	\expandafter\newcommand\csname #1check\endcsname{\textcolor{#1}{\ding{52}}}%
}
\colourcheck{blue}
\colourcheck{green}
\colourcheck{red}
\newtcolorbox{boxB}[2][]{%
  enhanced,colback=white,colframe=black,coltitle=black,
  sharp corners,
  toprule=1.0pt,
  rightrule=0.3pt,
  leftrule=0pt,
  bottomrule=0pt,
  fonttitle=\itshape\scshape\large,
  left=0pt,right=5pt,top=5pt,bottom=3pt,
  attach boxed title to top right={yshift=-0.3\baselineskip-0.4pt,xshift=-5mm},
  boxed title style={tile,size=minimal,left=0.2mm,right=0.5mm,
    colback=white,before upper=\strut},
  title=#2,#1
}

\newtcolorbox{outputbox}[1][]{ 
  colback=lightgray!10,  
  colframe=gray,      
  fontupper=\scriptsize,
  left=1mm, right=1mm, top=0.5mm, bottom=0.5mm,  
  before=\definecolor{royalblue}{rgb}{0.2549, 0.4118, 0.8824} 
}

\definecolor{lightred}{rgb}{1,0.85,0.85}
\definecolor{lightblue}{rgb}{0.85,0.92,1}
\definecolor{lightgreen}{rgb}{0.85,1,0.85}

\newtcolorbox{zoneboxred}{
  colback=lightred,
  colframe=red!50!red!50,
  boxrule=0pt,
  sharp corners,
  enhanced,
  left=1mm, right=1mm, top=0.5mm, bottom=0.5mm,
  before skip=1mm, after skip=1mm
}

\newtcolorbox{zoneboxblue}{
  colback=lightblue,
  colframe=blue!50!blue!50,
  boxrule=0pt,
  sharp corners,
  enhanced,
  left=1mm, right=1mm, top=0.5mm, bottom=0.5mm,
  before skip=1mm, after skip=1mm
}

\newtcolorbox{zoneboxgreen}{
  colback=lightgreen,
  colframe=green!50!green!50,B
  boxrule=0pt,
  sharp corners,
  enhanced,
  left=1mm, right=1mm, top=0.5mm, bottom=0.5mm,
  before skip=1mm, after skip=1mm
}

\newboolean{showcomments}
\setboolean{showcomments}{true}
\ifthenelse{\boolean{showcomments}}
 { \newcommand{\mynote}[2]{
      \fbox{\bfseries\sffamily\scriptsize#1}
        {\small$\blacktriangleright$\textsf{\emph{#2}}$\blacktriangleleft$}}}
        { \newcommand{\mynote}[2]{}}

\usepackage{tikz}

\newcolumntype{L}[1]{>{\raggedright\arraybackslash}p{#1}}

\newcommand{\code}[1]{{\footnotesize\texttt{#1}}}
\usepackage{amsthm}
\definecolor{dkgreen}{rgb}{0,0.6,0}
\definecolor{gray}{rgb}{0.5,0.5,0.5}
\definecolor{lightgray}{rgb}{211, 211, 211}
\definecolor{mauve}{rgb}{0.58,0,0.82}

\definecolor{custom-red}{rgb}{1,0,0}
\definecolor{custom-blue}{rgb}{0,0,0}

\definecolor{c1}{HTML}{f4cccc}
\definecolor{c2}{HTML}{f5cdcd}
\definecolor{c3}{HTML}{fffcfc}
\definecolor{c4}{HTML}{ffffff}
\definecolor{c5}{HTML}{ffffff}
\definecolor{c6}{HTML}{fffdfd}
\definecolor{c7}{HTML}{f5cfcf}
\definecolor{c8}{HTML}{fffbfb}
\definecolor{c9}{HTML}{ffffff}
\definecolor{c10}{HTML}{fffdfd}
\definecolor{c11}{HTML}{fefafa}
\definecolor{c12}{HTML}{fef7f7}
\definecolor{c13}{HTML}{ffffff}
\definecolor{c14}{HTML}{fffefe}
\definecolor{c15}{HTML}{ffffff}
\definecolor{c16}{HTML}{fefafa}
\definecolor{c17}{HTML}{fdf3f3}
\definecolor{c18}{HTML}{fffefe}
\definecolor{c19}{HTML}{fdf5f5}
\definecolor{c20}{HTML}{ffffff}

\definecolor{customcolor}{HTML}{CAEEFB}

\definecolor{dkgreen}{rgb}{0,0.6,0}
\definecolor{gray}{rgb}{0.5,0.5,0.5}
\definecolor{mauve}{rgb}{0.58,0,0.82}
\lstdefinestyle{textStyle}{
    keywordstyle= \color{blue!70},
    commentstyle= \color{red!50!green!50!blue!50},
    stringstyle=\color{red!70},
    frame=shadowbox,
    rulesepcolor= \color{red!20!green!20!blue!20} ,
    xleftmargin=1.5em,xrightmargin=0em, aboveskip=1em,
    framexleftmargin=1.5em,
            numbersep= 5pt,
    basicstyle=\scriptsize\ttfamily,
    numberstyle=\scriptsize\ttfamily,
    emphstyle=\bfseries,
    numbers=none
}

\definecolor{codegray}{gray}{0.5}
\definecolor{codepurple}{rgb}{0.58,0,0.82}
\definecolor{backcolour}{rgb}{0.95,0.95,0.92}
\definecolor{blue}{rgb}{0,0,1}

\lstdefinestyle{pythonStyle}{
    backgroundcolor=\color{backcolour},
    commentstyle=\color{codegray},
    keywordstyle=\color{magenta},
    numberstyle=\tiny\color{codegray},
    stringstyle=\color{codepurple},
    basicstyle=\ttfamily\footnotesize,
    breakatwhitespace=false,
    breaklines=true,
    captionpos=b,
    keepspaces=true,
    numbers=left,
    numbersep=5pt,
    showspaces=false,
    showstringspaces=false,
    showtabs=false,
    tabsize=2,
    language=Python,
    escapeinside={(*@}{@*)}
}

\lstdefinestyle{htmlStyle}{
    language=Python,
    numbers=left,
    numberstyle= \tiny,
    keywordstyle= \color{blue!70},
    commentstyle= \color{red!50!green!50!blue!50},
    stringstyle=\color{red!70},
    frame=shadowbox,
    rulesepcolor= \color{red!20!green!20!blue!20} ,
    xleftmargin=1.5em,xrightmargin=0em, aboveskip=1em,
    framexleftmargin=1.5em,
            numbersep= 5pt,
    basicstyle=\scriptsize\ttfamily,
    numberstyle=\scriptsize\ttfamily,
    emphstyle=\bfseries,
    morekeywords={doctype, html, head, body, title, script, div, span, a, data-result} 
}

\definecolor{eclipseStrings}{RGB}{42,0.0,255}
\definecolor{eclipseKeywords}{RGB}{127,0,85}
\colorlet{numb}{magenta!60!black}

\lstdefinelanguage{json}{
    basicstyle=\scriptsize\ttfamily,
    numberstyle=\scriptsize\ttfamily,
    emphstyle=\bfseries,
    commentstyle=\color{eclipseStrings}, 
    stringstyle=\color{eclipseKeywords}, 
    numbers=left,
    numberstyle= \tiny,
    keywordstyle= \color{blue!70},
    commentstyle= \color{red!50!green!50!blue!50},
    frame=shadowbox,
    rulesepcolor= \color{red!20!green!20!blue!20} ,
    xleftmargin=1.5em,xrightmargin=0em, aboveskip=1em,
    framexleftmargin=1.5em,
            numbersep= 5pt,
    showstringspaces=false,
    breaklines=true,
    string=[s]{"}{"},
    comment=[l]{:\ "},
    morecomment=[l]{:"},
    literate=
        *{0}{{{\color{numb}0}}}{1}
         {1}{{{\color{numb}1}}}{1}
         {2}{{{\color{numb}2}}}{1}
         {3}{{{\color{numb}3}}}{1}
         {4}{{{\color{numb}4}}}{1}
         {5}{{{\color{numb}5}}}{1}
         {6}{{{\color{numb}6}}}{1}
         {7}{{{\color{numb}7}}}{1}
         {8}{{{\color{numb}8}}}{1}
         {9}{{{\color{numb}9}}}{1}
}

\usepackage{tikz}

\definecolor{request}{HTML}{D80073}
\definecolor{response}{HTML}{0050EF}
\definecolor{circle_color}{HTML}{CCE5FF}

\usepackage{graphicx}
\usepackage{wrapfig}

\setcopyright{cc}
\setcctype{by-nc-nd}
\acmDOI{10.1145/3832202}
\acmYear{2026}
\acmJournal{PACMSE}
\acmVolume{3}
\acmNumber{ISSTA}
\acmArticle{ISSTA111}
\acmMonth{10}
\acmSubmissionID{issta26main-p985-p}
\received{2026-01-30}
\received[accepted]{2026-06-25}

\begin{document}





\title{On Behavioral Alignment of Model-Code and Human-Code Understandability via Behavioral Proxies}




\author{Xiaokai Rong}
\orcid{0009-0000-8457-8528}
\affiliation{%
  \institution{University of Texas at Dallas}
  \city{Dallas}
  \country{USA}
}
\email{xiaokai.rong@utdallas.edu}

\author{Aashish Yadavally}
\orcid{0000-0001-8785-6319}
\affiliation{%
  \institution{University of Central Florida}
  \city{Orlando}
  \country{USA}
}
\email{aashish.yadavally@ucf.edu}

\author{Hridya Dhulipala}
\orcid{0009-0001-4474-2984}
\affiliation{%
  \institution{University of Texas at Dallas}
  \city{Dallas}
  \country{USA}
}
\email{Hridya.Dhulipala@utdallas.edu}

\author{Anh H. N. Nguyen}
\orcid{0009-0006-9438-1888}
\affiliation{%
  \institution{University of Texas at Dallas}
  \city{Dallas}
  \country{USA}
}
\email{anguyen0210@gmail.com}

\author{Tien N. Nguyen}
\orcid{0009-0006-7962-6090}
\affiliation{%
  \institution{University of Texas at Dallas}
  \city{Dallas}
  \country{USA}
}
\email{tien.n.nguyen@utdallas.edu}

\begin{abstract}
Code understandability is a critical aspect of software quality. Prior research has largely focused on this attribute from a human-centric or code-centric perspective, while it should be viewed as a {\em relational property} arising from the interaction between a reader and the code. With the increasing adoption of large language models (LLMs) in software engineering, we posit that the notion of ``reader'' should be generalized to encompass both humans and models. Building on this relational perspective, we extend the concept of code understandability to distinguish between human and model code understandability, aiming to investigate the {\color{custom-blue}{behavioral alignment}} between the two. To this end, we use a dataset from a prior study containing human-rated judgments of understandability across diverse participant groups, and evaluate multiple open and closed-source LLMs.
To operationalize {\color{custom-blue}{such behavioral alignment}},
we introduce four {\color{custom-blue}{behavioral proxies of model-code understandability (BPMU)}}: P0, based on program comprehension-focused question answering; and P1--P3, based on program intent summarization (derived from our proposed semantic self-consistency).
Our findings show that LLMs exhibit stronger {\color{custom-blue}{behavioral alignment}} with general human code understandability than previously used shallow machine learning baselines. Stratified analyses further reveal that model code understandability aligns most closely with {\em professional developers}, but significantly less with other student groups. 
Role-conditioned prompting does not improve the alignment for the latter, suggesting that current LLMs cannot accurately mimic the perspectives of human readers with varying expertise. Finally, we demonstrate that semantic self-consistency is a reliable and extensible measure to be used as a behavioral proxy for quantifying model code understandability, with broad implications in both software engineering research and practice.

\end{abstract}

\ccsdesc[500]{Computing methodologies~Neural networks}
\ccsdesc[500]{Software and its engineering}

\keywords{AI4SE, Large Language Models, Model Understandability}





\maketitle

\section{Introduction}
\label{sec:intro}

Large Language Models (LLMs) have shown strong capability in several code understanding tasks.
As LLMs move from being occasional helpers to everyday collaborators in how developers read, review, and modify code, it becomes crucial {\em to understand whether the LLMs' ``understanding'' matches what humans actually experience}. A natural question is: {\bf if a code fragment is understandable (or non-understandable) to humans, is it also understandable (or non-understandable) to LLMs}? In other words, do human judgments of code understandability align with the model's ability to correctly interpret and reason about that~code?


The answer to this alignment question is important because it directly affects the reliability of LLM-assisted development. This alignment (or lack thereof) has deep implications. If model understanding fully or partially aligns with human understanding, it could enable more automation through model-in-the-loop processes, such as automated code reviews for human-maintained code repositories, LLM-generated documentation acceptable to developers, model-assisted naming suggestions, personalized developer tooling tailored to their expertise and objectives, and even empirical studies with LLMs in place of human participants.
In contrast, if model understanding does not align with human understanding, more human-in-the-loop collaboration will be required: LLM-generated code may be hard for human developers to maintain (even if syntactically correct, code that ``makes sense'' to models may be confusing to humans), hybrid human--AI code review process that combines LLM suggestions with human judgment might be needed, {\em etc}. 

To answer the question, we hypothesize that code understandability is viewed as a {\bf relational property} {\em arising from the interaction between a reader and the code}. With the increasing adoption of LLMs, we {\em generalize the notion of a ``reader'' to encompass both humans and models}. 
This distinction is important. For humans, each individual with diverse background and expertise
interprets the same program differently as Peitek {\em et al.}~\cite{peitek2022correlates} reported that professional programmers with high efficacy understand source code more targeted and with lower cognitive load.
LLMs too, either due to their training paradigms or prompt engineering, might reason about a program in~different~ways.~A relational view thus broadens the concept of code understandability (originally defined only for humans in program comprehension~\cite{BROOKS1983543,rajlich2002role,scalabrino2021automatically}) to {\bf human-code} and {\bf model-code understandability}.

This hypothesis also reflects our viewpoint that departs from the existing one where understandability is treated as a sole property of the program itself--something that can be derived from static syntactic or semantic measures (e.g., cyclomatic complexity, nesting depth, identifier statistics). Instead, under our perspective of understandability as relational, 
{\em the same fragment may be ``understandable'' to one interpreter and not to another, making alignment between human and LLM understandability an empirical question rather than an assumption implied by {\bf code-only viewpoint}. Our viewpoint is also {\em different from the {\bf human-centric perspective of programs} in which humans' data is used to train to teach ML models be close to humans' behaviors~\cite{scalabrino2021automatically,zhang2024eyetrans,wallace2025programmer}. 

In this work, we aim to {\bf investigate whether human-code understandability aligns with model-code understandability}, and how {\color{custom-blue}{behavioral proxies}} (or {\em proxies} for short) for both align with each other. We focus on the behavioral aspect of understandability, i.e., whether humans and models exhibit similar observable judgments and outputs when reasoning about code. 

We conducted a series of experiments to examine the {\color{custom-blue} behavioral alignment} between model code understandability and human code understandability (human understandability for short). 

{\bf Behavioral Proxies for Human-Code Understandability.} Scalabrino {\em et al.}~\cite{scalabrino2021automatically} 
conducted a study involving 63 human participants
who evaluated programs from open-source repositories, yielding a total of 444 assessments. 
The authors broadly measured two key {\color{custom-blue}behavioral proxies}: (a) {\em perceived understandability} (PBU), via developers' self-reporting, and (b) {\em actual understandability} (ABU), assessed through multiple-choice questions about program behaviors (Section~\ref{sec:proxies-hcu}). 


{\bf Behavioral Proxies for Model-Code Understandability.} In our study, we adopt this dataset and the above ABU metric as the proxies to measure human code understandability. To measure model code understandability, we first performed a formative preliminary study 
for our subject-based proxy P0 for model understandability. In the study, we treat the LLM as a participant in the empirical study, answering comprehensive, multiple-choice questions, which were reused from Scalabrino {\em et al.} to derive the proxy for actual understandability; but with the target LLM serving as participants. We found that using LLMs (particularly larger models) as subjects (P0) yields {\em partial alignment} with code understanding (ABU) of general human reader population and {\em strong alignment} with that of professional developers. However, this proxy P0 cannot be fully automated because it requires the creation of the multi-choice questions and answers for any new source code dataset.

To generalize our code understandability measure for models, we design a range of {\color{custom-blue}behavioral proxies of model code understandability (BPMU)} (P1--P3).
P1--P3 can be categorized as: {\em unsupervised} (P1) and {\em supervised} (P2--P3).
We focus on quantifying the ability of LLMs to produce consistent and accurate natural language summaries of a program's intent. We adopt this task since such descriptions explicitly capture the intended functionality, comprehending the program structure, control flow, and semantics to bridge higher-level program intent with lower-level implementation~\cite{DBLP:conf/icse/HaiducAM10,DBLP:journals/corr/abs-2107-07112}.

To quantify model code understandability (under proxies P1--P3), we introduce {\bf semantic self-consistency}. This builds on the principle of self-consistency in LLMs~\cite{Wang2023self}, where the reliability of a model's reasoning is judged by the degree to which multiple reasoning paths converge to consistent outcomes. The intuition is that if a model truly understands a program, it should generate multiple intent summaries for that program that are semantically consistent with each other across different reasoning paths. 
That is, {\em highly variable summaries suggest model confusion, while consistent ones indicate greater confidence and program understanding}.

In {\em unsupervised semantic self-consistency} (P1), we compute the mean pairwise semantic similarity among multiple summaries sampled from a model for a given program. Using a held-out calibration set, we derive a global similarity threshold to binarize scores for unseen programs into \textsc{High} (model certainty) and \textsc{Low} (model confusion). However, with P1, self-consistency can be \textsc{High} even when the summaries are mutually aligned but wrong. We refer to this as {\em high-confidence hallucinations}.


In contrast, the {\em supervised} variants of {\em semantic self-consistency} leverage external feedback from an expert LLM (chosen from a pool of state-of-the-art LLMs), thereby reducing the risk of such high-confidence hallucinations. In P2 (LLM-as-a-Meta-Reviewer), we incorporate this signal by comparing the target model's summaries to a reference summary generated by the expert LLM, with self-consistency being enforced by requiring the majority of summaries to meet the calibrated similarity threshold (derived as in P1).
In P3 (LLM-as-a-Judge), the expert LLM directly evaluates whether the target model's summaries capture the program's intent. Here, self-consistency is enforced by aggregating judgments across multiple summaries: if the expert LLM consistently renders positive evaluations ({\em i.e.}, a majority of summaries are adjudged correct), the model code understandability on the program is classified as \textsc{High}, otherwise as \textsc{Low}. By design, {\color{custom-blue}P1--P3 are {\bf automated BPMU}, i.e., {\em automated measures of model understandability}}, which can be extended to unseen programs. In contrast, P0 is tied to having access to {\em program comprehension-focused~questions}.


{\bf {\color{custom-blue}Behavioral Alignment} between P1--P3 and ABU.} 
In Scalabrino {\em et al.}~\cite{scalabrino2021automatically}'s {\em human-centric} view, the authors used ABU as measures of human understandability, treated these measures as ground truth, and trained various ML models to predict them. Their approach is thus explicitly {\bf human-centric}, whereas ours is not framed as predicting human judgments. We use PBU and ABU as behavioral proxies to quantify human-code understandability, enabling us to measure the behavioral alignment between P1--P3 on the model side and ABU on the human~side. We formulate the alignment question as a binary classification for LLMs via P0--P3 on the model side with the human-rated judgments of ABU serving as ground-truth labels on the human side.  



\subsection*{Summary of Key Findings:}




    (1) {\em LLM understandability aligns most closely with that of professional developers.} For undergraduates, master's, and even PhD students, the alignment between LLM predictions and human-rated judgments is notably weaker. These groups exhibit greater variability in how they judge understandability, likely due to differences in programming experience, language exposure, and sometimes, even over-confidence in understanding capabilities (e.g., in 19.3\% of the cases, bachelor's students assumed they understood a code snippet, but could only answer $\leq 1$ comprehension-focused question correctly). 
    In contrast, LLMs, which have largely been trained on open-source repositories, a good proportion of which originates from professional developers, seemed to internalize their comprehension patterns. This finding also supports our {\bf relational} view on reader-code understandability, which depends on the interaction between human or LLM readers and source~code.



    (2) {\em Role-conditioned prompting does not bridge gaps in developer expertise.} Our results show that prompting LLMs with role designations to enable them to mimic the perspectives of different reader groups does not work, and the alignment between model code understandability and human judgments do not reliably improve (or sometimes drop). This suggests that the models fall back on surface-level heuristics ({\em e.g.}, simpler wording for bachelor's students), without meaningfully capturing the distinct comprehension styles of diverse human readers. Future work may need to explore few-shot learning,  targeted fine-tuning or post-training strategies to induce such~behaviors.


    (3) While unsupervised semantic self-consistency (P1) captures {\em partial alignment} with human readers, it fails to account for high-confidence hallucinations. Incorporating external feedback in the form of reference summaries (P2) or direct judgments (P3) 
    addresses this limitation, yielding {\em partial alignment} with code understandability of general human reader population.

    (4) By comparing with the subject-based proxy (P0) via Q\&As, which is the closest approximation to {\em actual model comprehension}, we found that the design of calibrated semantic self-consistency P1--P3 provides a reliable method for quantifying model code understandability. By design, it is also extensible to unseen programs, making it a valuable tool for future research. 


(5) In Section~\ref{sec:implications}, we expand on the implications of our findings on SE research and practice. 


\section{{\color{custom-blue}Behavioral Proxies} for Human Code Understandability}\label{sec:proxies-hcu}


Scalabrino {\em et al.}~\cite{scalabrino2021automatically} conducted an empirical study on human code understandability with 63 participants, including professional developers as well as undergraduate, master's and Ph.D. students. The study produced 444 evaluations of Java code snippets, in which participants first judged whether they understood a snippet and then answered verification questions. From these, the authors defined behavioral proxies to operationalize human judgments of code understandability: 


\subsubsection*{\bf Perceived Binary Understandability (PBU)}
After reading a code snippet, each participant declared ``{\em I cannot understand the method}'' or ``{\em I understood the method}''. This measure is subjective and self-reported, reflecting perceived rather than demonstrated comprehension. Formally, given a program $P \in \mathcal{D}$ and a participant $h$, let $y_h(P) \in \{\textsc{False},\textsc{True}\}$ denote the self-reported label. Then, 
\[
\text{PBU}(P) = 
\begin{cases}
1, & \text{if } y_h(P) = \textsc{True}, \\[6pt]
0, & \text{otherwise}.
\end{cases}
\]
\subsubsection*{\bf Actual Understandability (AU)}
To capture actual comprehension, human participants were asked a set of $q$ verification questions about each snippet. AU captures the proportion of correct answers provided by the participants. 
This contiguous score represents a degree of understanding, capturing beyond binary perception.
Formally, given a program $P \in \mathcal{D}$, let $\{a^*_1, a^*_2, \dots, a^*_q\}$ denote the gold answers and $\{a^h_1, a^h_2, \dots, a^h_q\}$ be the participants' responses. Then,
\begin{equation*}
\text{AU}(P) = \frac{1}{q}\sum_{i=1}^q \mathbf{1}\{a^h_i = a^*_i\}
\end{equation*}
\subsubsection*{\bf Actual Binary Understandability (ABU$_{k\%}$)}
ABU$_{k\%}$ is defined as a binarized variant of AU obtained by applying a threshold of $k\%$ over AU. Formally, for a program $P \in \mathcal{D}$ with actual understandability score $\text{AU}(P)$:
\[
\text{ABU}(P) = 
\begin{cases}
1, & \text{if } \text{AU}(P) \geq k\%, \\[6pt]
0, & \text{otherwise}.
\end{cases}
\]

In addition to these proxies for {\em correctness}, the authors considered the {\em time} to understand code snippets. Due to our focus on the alignment between model and human code understandability, we only adopt ABU$_{k\%}$ as they are the most reliable ground-truth labels of actual human judgment~\cite{scalabrino2021automatically}.

\section{Preliminary Study: (Subject-Based) Proxy P0: LLM-as-a-Subject}
\label{sec:prelim}



In Scalabrino {\em et al.}~\cite{scalabrino2021automatically}'s study, they interviewed developers with questions-and-answers to evaluate their understanding of code snippets. Building on this, we use LLMs as subjects instead of human participants, effectively running an LLM subject-based empirical study. {\em Models must directly answer comprehension questions on code snippets, thereby providing insights into model code understandability}.
Formally, for each program $P \in \mathcal{D}$ we pose a set of $q$ comprehension-focused multiple-choice questions (each having four choices), with gold answers from a human subject-based study denoted by $\{a^*_1, a^*_2, \dots, a^*_q\}$. To obtain robust responses from the target models, we adopt self-consistency; {\em i.e.}, each question $q_i$ is prompted $r$ times under stochastic decoding, producing a set of answers $\{a_i^{(1)}, a_i^{(2)} \dots, a_i^{(r)}\}$. These are then aggregated by majority voting to yield a final answer $\hat{a}_i$. The {\em subject-based understandability score} of a model with regard to a program $P$ is defined as:
\begin{equation*}
\mathrm{SUBJ}(P) = \frac{1}{q}\sum_{i=1}^{q} \mathbf{1}\{\hat{a}_i = a^*_i\}
\end{equation*}
A higher $\mathrm{SUBJ}(P)$ indicates that the model $\mathcal{M}$ correctly and consistently answers comprehension questions, reflecting stronger model code understandability. To classify these scores, we adopt the methodology used to derive $ABU_{p\%}$ (see Section~\ref{sec:proxies-hcu}), where $p \in [0, 1]$ is the correctness threshold. A program $P$ is labeled \textsc{HIGH} if $\mathrm{SUBJ}(P) \geq p$, and \textsc{LOW} otherwise. 
\[
\mathrm{Understandability}(P) = 
\begin{cases}
\textsc{High}, & \text{if } \mathrm{SUBJ}(P) \geq p, \\[6pt]
\textsc{Low}, & \text{otherwise}.
\end{cases}
\]
We will present our result on the alignment of the subject-based proxy P0 for LLMs and the proxy APU for humans~\cite{scalabrino2021automatically} in Section~\ref{sec:qanda-results}.

\section{{\color{custom-blue}{Behavioral Proxies of Model Code Understandability (BPMU)}}}\label{sec:proxies-mcu}

In the subject-based proxy P0, LLMs can simulate question answering, but P0 is not fully automated because each new source-code dataset requires newly crafted multiple-choice Q\&As.
Thus, toward a generalized measure of how LLMs understand code, we propose two classes of 
{\color{custom-blue} behavioral~proxies of model code understandability (BPMU)}:
{\em unsupervised} and {\em supervised}. These proxies differ in the source of the signal used to evaluate understandability. Both unsupervised and supervised proxies rely on the model's internal behavior, quantifying its understanding via {\em semantic~self-consistency}, with the latter grounding the model's outputs against an external signal from an `expert' model. 


\subsection{(Unsupervised) Proxy P1: Semantic Self-Consistency}\label{sec:p1}
\subsubsection{Semantic Self-Consistency (SSC)}
Natural language summaries of a given code snippet capture developer intent by explicitly describing its intended functionality, bridging the gap between its higher-level understanding and lower-level implementation details~\cite{DBLP:conf/icse/HaiducAM10,DBLP:journals/corr/abs-2107-07112}. Building on this, we propose to use an LLM's ability to consistently summarize the code intent as an automated
{\color{custom-blue}BPMU}.
If the model produces semantically similar intent summaries across diverse thought paths, it suggests a stable internal representation of the code and thus higher understandability. In contrast, deliberate thinking and a greater divergence across summaries indicates lower understandability.

This intuition draws parallels from the idea of self-consistency~\cite{Wang2023self}, which adopts a ``sample-and-marginalize'' approach to improve reliability of model outputs by converging on the most consistent answer. In contrast, our goal is not to marginalize, but to quantify the semantic alignment among the model's outputs. Therefore, we refer to this proxy as {\bf semantic self-consistency} (SSC), {\em reflecting the degree to which the model's generated intent summaries align in meaning across diverse samples}. Formally, given $k$ summaries $\mathcal{S} = \{s_1, s_2, \dots, s_k\}$ produced by a model $\mathcal{M}$, we define the {\em semantic self-consistency score} as the mean pairwise similarity:
\begin{equation}
    \text{SSC}(\mathcal{S}) = \frac{2}{k(k-1)} \sum_{1\leq i < j \leq k} M(s_i, s_j).
\end{equation}
where $M(\cdot, \cdot)$ denotes a similarity function between two summaries. In this work, we instantiate $M$ with BERTScore-F1~\cite{zhang2019bertscore}, which measures semantic similarity in the embedding space of a pre-trained language model. Specifically, we adopt DeBERTa-base~\cite{DBLP:journals/corr/abs-2006-03654} as the underlying encoder. 

\subsubsection{Calibrating Semantic Self-Consistency of Intent Summaries}\label{sec:p1-calibration}
While semantic self-consistency (SSC) provides a continuous measure of alignment, interpreting this score requires a notion of confidence. Thus, we adopt a calibration scheme for a given dataset $\mathcal{D}$ of code snippets. 
For a program $P\in\mathcal{D}$, we first partition the LLM-generated summaries $\mathcal{S}$ into a calibration subset $\mathcal{S}\textsubscript{cal} = \{s_1, s_2, \dots, s_{k_i}\}$ and test subset $\mathcal{S}\textsubscript{test} = \{s_{k_i+1}, s_{k_i+2}, \dots, s_{k}\}$. We then compute the semantic self-consistency score for the calibration subset, $\text{SSC}(\mathcal{S}\textsubscript{cal})$, and define the {\em nonconformity score} as 
\begin{equation*}
\alpha = 1 - \text{SSC}(\mathcal{S}\textsubscript{cal})
\end{equation*}
In general, $\alpha$ quantifies the extent to which the generated summaries fail to conform to one another: a lower $\alpha$ reflects stronger model agreement ({\em i.e.}, the summaries are semantically consistent), whereas a higher $\alpha$ indicates greater variability, signaling potential model confusion.

\vspace{1pt}
{\em Setting the Global Threshold from Summary Calibration Subsets.}
Let $\alpha^{(1)}, \alpha^{(2)}, \dots, \alpha^{(|\mathcal{D}|)}$ be the {\em calibration nonconformity scores} for all programs in $\mathcal{D}$. Given a desired coverage level $1 - \epsilon$ ({\em e.g.}, $0.9$), we compute a global threshold as the $(1-\epsilon)$ empirical quantile of these scores. Formally,
\begin{equation*}
\text{Global threshold}, \tau = Q_{1-\epsilon} \big(\{\alpha^{(1)}, \alpha^{(2)}, \dots, \alpha^{(|\mathcal{D}|)}\}\big) 
\end{equation*}
where $Q_p(.)$ denotes the $p$th-quantile. In other words, $\tau$ is the cutoff such that at least ($1-\epsilon$) of the programs achieve semantic self-consistency above this threshold, while the least consistent $\epsilon$-fraction are tolerated as exceptions ({\em i.e.}, $\epsilon$ directly controls how tolerant the criterion is).


\vspace{1pt}
{\em Classification of Summary Test Subsets}.
For each test subset $\mathcal{S}\textsubscript{test}$ of model-generated summaries for a program $P \in \mathcal{D}$, let $\alpha_t$ denote its corresponding nonconformity score. The model code understandability of the program $P$ is then classified as \textsc{High} or \textsc{Low} through P1 as follows:
\[
\mathrm{Understandability}(P) =
\begin{cases}
\textsc{High} \text{ (\textit{$\mathcal{M}$ exhibits certainty})}, & \alpha_t \leq \tau, \\
\textsc{Low} \text{ (\textit{$\mathcal{M}$ exhibits confusion})}, & \alpha_t > \tau.
\end{cases}
\]

\subsection{(Supervised) Proxy P2: Semantic Self-Consistency with LLM-as-a-Meta-Reviewer}\label{sec:p2}
While semantic self-consistency (as defined in Section~\ref{sec:p1}) quantifies the internal agreement of a model's generated summaries, it does not provide correctness guarantees. For instance, a model may consistently produce summaries that are semantically aligned with one another but nonetheless might be incorrect and do not capture the intent of the given code snippet. To this end, we propose a {\em reference-aware variant of semantic self-consistency}, where the model's outputs are grounded against a reference intent summary produced by an expert LLM.
That is, {\color{custom-blue}{according to this supervised, reference-aware BPMU}}, a model is considered to ``understand'' the given snippet if its summaries are both internally ``consistent'' and ``close'' to the reference summary from an expert LLM. 


\subsubsection{Generating the Reference Intent Summary}
We employ three state-of-the-art LLMs: {\color{custom-blue}GPT-5}, Claude Opus 4.1, and Gemini-2.5 Pro as meta-reviewers. For each program $P \in \mathcal{D}$, we first generate $m$ summaries from each meta-reviewer. Following the procedure described in Section~\ref{sec:p1-calibration}, we then retain the top-$n$ summaries ($n<m$) adjudged to be the most confident for each meta-reviewer, resulting in a pool $\mathcal{R}$ of high-quality reference summaries ($|\mathcal{R}|=3n$). From this pool, we randomly select one $r^* \in \mathcal{R}$ to serve as the reference summary for grounding the target model's outputs.

\subsubsection{Calibrating Semantic Self-Consistency of Intent Summaries Against a Reference}
Following the self-calibration step used to select the reference summary, we reuse the calibration and test subsets from Section~\ref{sec:p1-calibration} ($\mathcal{S}\textsubscript{cal}$ and $\mathcal{S}\textsubscript{test}$, respectively) for each program $P \in \mathcal{D}$. The {\em reference-aware semantic self-consistency score} is then defined as the mean semantic similarity between each model-generated summary and the selected reference summary $r^*$:
\begin{equation}
    \text{SSC}\textsuperscript{meta}(\mathcal{S}_x, r^*) = \frac{1}{|S_x|}\sum_{i=1}^{|\mathcal{S}_x|} M(s_i, r^*)
\end{equation}
where $M(\cdot,\cdot)$ again denotes a semantic similarity function (instantiated with BERTScore-F1),
and $\mathcal{S}_x$ denotes either the calibration or test subsets. A higher SSC\textsuperscript{meta} score indicates that the model~$\mathcal{M}$ {\em not only generates intent summaries that are internally consistent but also semantically aligned~with a trusted external reference}, signaling stronger understanding. We define the nonconformity score as
\begin{equation*}
\alpha = 1 - \text{SSC}\textsuperscript{meta}(\mathcal{S}_x, r^*).
\end{equation*} 
Here, a lower $\alpha$ indicates stronger self-consistency and alignment to the reference ({\em greater confidence}), while a higher $\alpha$ suggests greater deviation from the meta-reviewer ({\em model confusion}).


\subsubsection{Setting Global Threshold and Classification of Summary Test Subsets}

We follow the procedure as in Section~\ref{sec:p1-calibration} to compute the global threshold $\tau$ for the calibration subset $\mathcal{S}\textsubscript{cal}$ as the $(1-\epsilon)$ empirical quantile of their corresponding calibration nonconformity scores. We adopt the same classification scheme to classify the model code understandability for each test subset $\mathcal{S}\textsubscript{test}$. Those with a nonconformity score $\alpha_t \leq \tau$ are labeled \textsc{High} ({\em i.e.}, the model's summaries are consistent and align closely with the reference), and with $\alpha_t > \tau$ are labeled \textsc{Low} ({\em i.e.}, diverge from the reference).

\subsection{(Supervised) Proxy P3: Semantic Self-Consistency with LLM-as-a-Judge}
The proxy P2 grounds model outputs against an expert-generated reference summary. Alternatively, we use another proxy where expert LLMs act directly as {\em judges}, leveraging their internal knowledge to {\em evaluate whether the summary from a target model captures the underlying program intent}.

\subsubsection{LLM-as-a-Judge}
We reuse the same pool of expert LLMs from Section~\ref{sec:p2}: {\color{custom-blue}GPT-5}, Claude Opus 4.1, and Gemini-2.5 Pro. For each program $P \in \mathcal{D}$, we randomly select one model from this pool (denoted by $\mathcal{M}^{*}_{\text{expert}}$) to serve as a judge. This expert then directly evaluates whether each summary $s_i \in \mathcal{S}$ generated by the target model correctly reflects the program's functionality and intent. Notably, unlike P2, this judgment relies on the expert's knowledge.
To improve the robustness of the judgment, we incorporate a self-consistency step before aggregation. Formally, for each summary $s_i \in \mathcal{S}$, the expert model $\mathcal{M}^{*}_{\text{expert}}$ is prompted multiple times under stochastic decoding, yielding a set of judgments $\{J^{(1)}(s_i), J^{(2)}(s_i), \dots, J^{(r)}(s_i)\}$, where each $J^{(j)}(s_i) \in \{\textsc{Correct}, \textsc{Incorrect}\}$. The final decision for summary $s_i$ is then obtained by majority voting across these sampled judgments: 

\[
\hat{J}(s_i) = 
\begin{cases}
\textsc{Correct}, & \text{if } \sum_{j=1}^{r} \mathbf{1}\{J^{(j)}(s_i) = \textsc{Correct}\} \geq \tfrac{r}{2}, \\[6pt]
\textsc{Incorrect}, & \text{otherwise}.
\end{cases}
\]
\subsubsection{Majority Voting over Intent Summaries}
For a given program $P \in \mathcal{D}$, after obtaining binary decisions $\mathcal{J} =\{\hat{J}(s_1), \hat{J}(s_2), \dots, \hat{J}(s_{k})\}$ for each $s_i \in \mathcal{S}$, semantic self-consistency score is determined by majority voting across all generated intent summaries:
\begin{equation}
    \text{SSC}\textsuperscript{judge}(\mathcal{S}, \mathcal{J}) = \frac{1}{k}\sum_{i=1}^{k} \mathbf{1}\{\hat{J}(s_i) = \textsc{Correct}\}
\end{equation}
In this proxy P3, model code understandability of the program is labeled \textsc{High} when the majority of its summaries are judged to correctly reflect the code's functionality and intent, and \textsc{Low} otherwise.
\[
\mathrm{Understandability}(P) = 
\begin{cases}
\textsc{High}, & \text{if } \text{SSC}\textsuperscript{judge}(\mathcal{S}, \mathcal{J}) \geq 0.5, \\[6pt]
\textsc{Low}, & \text{otherwise}.
\end{cases}
\]

\section{Empirical Study Design}\label{sec:design}
\subsection{Research Questions}

{\color{custom-blue}Given a set of programs $\mathcal{D}$, we aim to examine whether the behavioral proxies of model code understandability (BPMU) (defined in Section~\ref{sec:proxies-mcu}) align with those of human code understandability in (defined in Section~\ref{sec:proxies-hcu})}. 
We formulate the following research questions:

\begin{enumerate}[label={\bf RQ\arabic*.}]
    
    \item Does model code understandability and human code understandability align via proxies?
    \begin{enumerate}[label=1.\arabic*,topsep=0pt, itemsep=0pt, leftmargin=-8pt, labelsep=3pt]

        \item ({\em Subject-Based}) [Preliminary Study] How well does model understandability measured by P0 (simulating human participants with LLMs as subjects) align with human readers?
        
        \item ({\em Unsupervised}) How well does model understandability from P1 align with human~readers?
        
        \item ({\em Supervised}) How well does model code understandability measured by semantic self-consistency using LLM-as-a-Meta-Reviewer (P2) and LLM-as-a-Judge (P3) for external feedback align with human readers?

        
    \end{enumerate}    
    \item ({\em Stratified Evaluation}) How well do our behavioral proxies of model code understandability (P1--P3) capture the perspectives of {\em different developer sub-populations} with varying expertise levels (from undergraduate and graduate students to professional developers)? 
    \item ({\em Role-Specific Conditioning}) Can large language models mimic the perspectives of different developer sub-populations when prompted with designated roles?
    \item How reliable is semantic self-consistency in capturing model code understandability?
\end{enumerate}

\subsection{Methodology}
\subsubsection{Experimental Dataset}
In our preliminary study, we evaluate whether the subject-based proxy P0 for model-code understandability align with the developers' annotations in Scalabrino {\em et al.}~\cite{scalabrino2021automatically}'s study.
In a later study, we evaluate whether the P1--P3  proxies for {\em model code understandability} (Section~\ref{sec:proxies-mcu}) align with the developers'. Accordingly, we focus on the human-centered proxy that provides ground-truth signals: 
{\em Actual Binary Understandability} (ABU\textsubscript{50\%}). 
ABU\textsubscript{50\%} (or ABU for short) is derived from how many functional comprehension~questions the subjects could answer correctly about the same programs (at least half correct indicating \textsc{True}).
In their study, among all human-rated evaluations, in 69.5\% of the cases, the developers perceived a program as {\em understandable} while only 50.2\% actually did. Therefore, we used only ABU\textsubscript{50\%}. When aggregated at the program level, these yield the ABU labels, exhibiting class imbalance with 82\% marked as {\em understandable}.

\subsubsection{Models}
We evaluated a set of competitive open-source/weight and closed LLMs in code summarization to ensure diversity across training paradigms and model sizes ({\em i.e.}, number of parameters). In particular, we include {\color{custom-blue}Phi-4-Reasoning, Qwen2.5-14B, Qwen3-14B} ~\cite{qwen2025qwen25technicalreport}, WizardCoder-15B~\cite{luo2025wizardcoderempoweringcodelarge}, StarCoder2-15B~\cite{lozhkov2024starcoder2stackv2}, and CodeLlama-13B~\cite{code_llama} to represent publicly available instruction-tuned LLMs for coding tasks. Among the proprietary models, we used GPT-3.5, GPT-4 and {\color{custom-blue}GPT-4o}~\cite{ChatGPT}.

\vspace{3pt}
{\em Expert LLMs.} We designated a pool of expert models to provide external feedback in both the LLM-as-a-Meta-Reviewer (P2) and LLM-as-a-Judge (P3) proxies. For this role, we use the state-of-the-art LLMs which show a strong performance on code understanding tasks including {\color{custom-blue}GPT-5}~\cite{ChatGPT}, Claude Opus 4.1~\cite{claude_opus41_2025}, and Gemini-2.5-Pro~\cite{gemini25_2025}.

\subsubsection{Evaluation Metrics} 
To assess the alignment of model code understandability with human code understandability, we 
formulate the problem as a binary classification task, i.e., whether a target model's answers on the (non-)understandability of programs via P0--P3 proxies align with that from humans on the same programs. We adopt standard binary classification metrics: {\em Accuracy} = $\frac{TP + TN}{TP + TN+ FP + FN}$, {\em Precision} = $\frac{TP}{TP+FP}$, {\em Recall} = $\frac{TP}{TP + FN}$, and {\em F1-Score} = $\frac{2 * Precision * Recall}{Precision + Recall}$. Here, $TP$, $TN$, $FP$, and $FN$ denote the number of true positives, true negatives, false positives, and false negatives, respectively, with respect to human understandability ratings. We also report {\em Area Under ROC Curve} (AUC), which provides a threshold-independent measure of discriminative ability. A higher AUC indicates that a proxy more effectively separates \textsc{High} and \textsc{Low} understandability cases. 
Finally, to mitigate the effects of class imbalance, we use F1-score as the primary measure of alignment.



\subsubsection{Procedure}\label{sec:implementation}
For all programs, each model generated 50 intent summaries per program. In the LLM-as-subject proxy P0, we assess comprehension by prompting the target LLM to answer 3 multiple-choice questions per program, each having 4 options~\cite{scalabrino2021automatically}. To ensure robustness through self-consistency, we obtain 50 responses per question, adopting majority voting to determine the final answer. Moreover, to mitigate any positional biases in option selection, we rotate the placement of the correct answer in each question across 4 independent runs and aggregate the responses.

For self-consistency P1--P3 (both (un)supervised), we randomly partitioned these into 25 summaries for calibration and 25 for testing. We computed the global threshold as the 90$^{th}$ empirical quartile of the calibration nonconformity scores (significance level $\epsilon$=0.1). This ensures that at least 90\% of the calibration programs exhibit nonconformity scores below the threshold (Section~\ref{sec:p1-calibration}).

{\color{custom-blue}In both supervised proxies P2 and P3, we use a pool of 3 above expert LLMs as a meta-reviewer or a judge to provide external feedback. During self-calibration of the expert models in the first setting, each generates 30 intent summaries per program. From these, we retain the Top-3, yielding a meta-reviewer reference pool of 9 summaries, and randomly select one of them to serve as the~gold reference summary. In the second setting, we randomly assign one expert model per program, prompting it to assess the alignment between the target model-generated intent summary and the given program 50 times ({\em i.e.}, majority voting is applied over $r=50$ judgments).}

\section{Assessing Behavioral Alignment between Model Code Understandability and Human Code Understandability (RQ1)}


\subsection{Preliminary Study (Subject-Based): Comparing \underline{LLMs as Subjects} and Humans in Code Understandability-Focused  Studies (P0)}~\label{sec:qanda-results}

Scalabrino {\em et al.}~\cite{scalabrino2021automatically} obtained~a reliable measure of human code understandability by asking human subjects multiple-choice verification questions. In this experiment, we replicate the procedure with LLMs in place of human participants, effectively treating the models as subjects. In Table~\ref{tab:qanda} ({\em left}), we present the results for this study. 

\begin{wraptable}{r}{0.5\textwidth} 
    \caption{Comparison of LLM-as-a-Subject (P0) with actual binary understandability (ABU) of human readers.}
    \vspace{-6pt}
    \footnotesize
    \centering
    \begin{tabular}{c|c|c|c|c|c}
    \toprule
    \multirow{2}{*}{\textbf{$\langle$Model, Code$\rangle$}} & \multicolumn{5}{c}{\textit{Evaluation Metrics} (\textit{in \%})} \\ \cline{2-6} 
                                          & A      & P      & R      & F1        & AUC     \\ \midrule
    Qwen2.5-14B                           & 57.6       & 65.7       &   53.5     & 66.7      &    58.3     \\
    WizardCoder-15B                       &  53.0      &  60.3    &  51.2   & 55.3      &    53.3     \\
    StarCoder2-15B                         &  44.4      & 56.3       &  10.5      & 17.6      & 49.8        \\
    CodeLlama-13B                         &    48.3    &    78.6    &   12.8     & 22.0      &    54.1     \\
        {\color{custom-blue}Qwen3-14B}                    & 57.0   & 57.2   & 96.5   & 71.9  & 56.3    \\
    {\color{custom-blue}Phi-4}                         & 57.6   & 57.6   & 96.5   & {\bf 72.2}  & 58.4    \\ \hline

    GPT-3.5                               & 60.3       & 59.7       & 93.0       & {\bf 72.7}      & 54.9        \\
    GPT-4                                 & 54.9       & 58.2       &  74.4      & 65.3      & 46.6        \\
    {\color{custom-blue}GPT-4o} & 57.6 & 57.6 & 96.5 & 72.2 & 58.6 \\

    \bottomrule
    \end{tabular}
    \label{tab:qanda}
\end{wraptable}

Among the smaller open-source models, we can see that {\color{custom-blue}Qwen3-14B and Phi-4-Reasoning achieve the highest F1-scores of 71.9\% and 72.2\%}, while StarCoder-15B and CodeLlama-13B perform poorly, with F1-scores of only 17.6\% and 22.0\%, respectively. The closed-source models perform consistently higher than open-source ones except Qwen3-14B and Phi4. We observe a counter-intuitive result: GPT-3.5 slightly performs better than GPT-4 and GPT-4o by 7.4\% and 0.5\% in F1-score. This trend is also similar with P1--P3, where they yield comparable alignment (Sections~\ref{sec:unsupervised-results},~\ref{sec:supervised-results}). 

A possible explanation is that GPT-4 and GPT-4o exhibit relatively stronger positional or option-selection bias in the multiple-choice format. As in Section~\ref{sec:implementation}, we rotated the correct answer among all options to mitigate such effects. 
Nevertheless, GPT-4's selections followed a less balanced distribution of 28.3\%--26.2\%--21.6\%--23.9\%, compared to 26.9\%--24.2\%--24.1\%--24.8\% for GPT-3.5. 
Combined with majority voting over multiple samples, this imbalance may have affected GPT-4's performance, despite its superior reasoning.

Overall, we found that with P0, model understandability has {\em partial alignment} with ABU, particu\-larly for larger models. Importantly, P0 shows the promise of treating LLMs as subjects in~comprehen\-sion-focused studies, while calling for evaluation schemes that are robust to format-specific biases.


\begin{tcolorbox}[top=2pt, bottom=2pt, left=5pt, right=5pt]
{\bf \em Findings (P0):} 
\begin{enumerate}[topsep=0pt, itemsep=0pt, start=1, leftmargin=*]
    \item Using LLMs as subjects (P0) yields {\color{custom-blue}{\em partial behavioral alignment}} with code understanding of general human reader population. However, such settings are prone to positional or option-selection biases in the multiple-choice format. This proxy cannot be fully automated.
\end{enumerate}
\end{tcolorbox}

\subsection{Unsupervised: Semantic Self-Consistency (P1)}~\label{sec:unsupervised-results}

\vspace{-9pt}
\subsubsection{Alignment of Unsupervised Proxy P1 and ABU} In Table~\ref{tab:unsupervised}, we report the alignment of P1 with the proxy for human code understandability via ABU. Across both public/weight-available and proprietary LLMs, for the open-source models, ABU F1-scores range from 65.7\% to 72.3\%. Closed-source models follow a similar trend, with ABU F1-scores of 71.1\%--71.9\%. 


Our further inspection revealed a crucial aspect. While P1 captures the effect of model confusion, it does not account for {\em high-confidence hallucinations}: a model generates fluent and consistent yet incorrect descriptions of program intent. In these cases, P1 captures an {\em illusion} of understanding even when true comprehension is missing. This motivated our supervised proxies P2-P3 (Section~\ref{sec:supervised-results}).

\begin{table}[t]
\footnotesize
\centering
\caption{Comparison of P1 (Semantic Self-Consistency) against 1) humans' Actual Binary Understandability (ABU) to assess {\bf behavioral alignment}, and 2) P0 (LLM-as-a-Subject) to assess {\bf cross-proxy reliability}. Values reported are Accuracy (A), Precision (P), Recall (R), F1, and AUC (\%). For P1 vs P0, only F1 is shown.}
\vspace{-6pt}
\begin{tabular}{l|c|c|c|c|c||c}
\toprule
\multicolumn{1}{c|}{\textbf{Model}}
& \multicolumn{6}{c}{\textit{Evaluation Metrics} (\textit{in} \%)} \\ \cline{2-7}
\multicolumn{1}{c|}{}
& \multicolumn{5}{c||}{\textbf{P1 (Sem. Self-Consistency) vs ABU}}
& \textbf{P1 vs P0} \\ \cline{2-7}
\multicolumn{1}{c|}{}
& A & P & R & F1 & AUC & F1 \\ \midrule

Qwen2.5-14B     & 56.9 & 58.7 & 82.6 & 68.6 & 52.8 & 65.3 \\
WizardCoder-15B & 58.3 & 58.4 & 93.0 & 71.7 & 52.7 & 58.2 \\
StarCoder2-15B   & 52.3 & 54.9 & 90.7 & 68.4 & 46.1 & 19.2 \\
CodeLlama-13B   & 53.6 & 56.8 & 77.9 & 65.7 & 49.7 & 14.6 \\ 
{\color{custom-blue}Qwen3-14B}   & 57.0 & 57.0 & 98.8 & \textbf{72.3} & 50.2 & 97.3 \\
{\color{custom-blue}Phi-4}   & 57.0 & 57.6 & 93.0 & 71.1 & 51.1 & 93.3 \\
\hline
GPT-3.5         & 56.3 & 56.8 & 97.7 & 71.8 & 49.6 & 92.9 \\
GPT-4           & 56.9 & 57.6 & 93.0 & 71.1 & 51.1 & 92.1 \\
{\color{custom-blue}GPT-4o} & 57.0 & 57.2 & 96.5 & {\bf 71.9} & 50.6 & 95.5 \\

\bottomrule
\end{tabular}
\label{tab:unsupervised}
\end{table}

Notably, P1 outperforms shallow machine learning models in Scalabrino {\em et al.}~\cite{scalabrino2021automatically}, which directly predicted human code understandability with F1-scores of 60.0\%--66.0\%. 
In fact, we observe that the LLMs can learn to model such complex relationships, with the use of multiple sampled intent summaries (as in P1) helping to account for the variability in human understanding.

\subsubsection{Reliability of Semantic Self-Consistency Proxy P1 for Model Code Understanding} In Table~\ref{tab:unsupervised}, we also show the cross-proxy reliability when we compare the unsupervised proxy P1 to the subject-based proxy P0. By design, our proxy with LLM-as-a-Subject (P0) involves target models answering the same questions as human participants for each program. As a result, P0 is most directly analogous to ABU and serves as the closest approximation of ``actual'' model code understandability.~To assess the reliability of 
P1, we compare it against P0. This provides a basis for evaluating the {\em extensibility} of P1 to settings where P0 is not applicable, i.e., comprehension-focused questions are not available.

As noted earlier, the poor comprehension of weaker open-source models such as StarCoder2-15B and CodeLlama-13B (F1-scores of 17.6\% and 22.0\%, respectively, in Table~\ref{tab:qanda}) translates to weak cross-proxy alignment between P1 and P0 (F1-scores of 0.0--19.2\%). {\color{custom-blue}In contrast, GPT-3.5, GPT-4, GPT-4o, Qwen3-14B, and Phi-4 demonstrate much higher alignment between P1 and P0 (F1-scores of 92.1\%--97.3\%).} Since P0 is evaluated in a question-and-answer setting, {\em simple memorization} of $\langle${\em program}, {\em intent summary}$\rangle$ pairs during model pretraining {\em would not by itself yield high correlation between P1 and P0}. Instead, this alignment arises from the design of calibrated semantic self-consistency, reinforcing both the reliability of our proxy P1 and its potential applicability to new datasets.

Overall, these findings show both the utility and limitations of our unsupervised proxy. While P1 provides evidence of moderate alignment between model and human understandability across diverse expertise, it remains incomplete. Semantic self-consistency should be interpreted as a complementary signal rather than a standalone proxy for model comprehension. We include a sensitivity analysis of thresholding effects on empirical coverage guarantees on the project~website~\cite{code-understanding}.

\begin{tcolorbox}[top=2pt, bottom=2pt, left=5pt, right=5pt]
{\bf \em Findings (P1):} 
\begin{enumerate}[topsep=0pt, itemsep=0pt, start=2, leftmargin=*]
    \item Via the unsupervised proxy P1, models show {\em partial {\color{custom-blue}behavioral alignment}} with general reader population but is insufficient, as it fails to account for high-confidence hallucinations.
\end{enumerate}
\end{tcolorbox}


\subsection{Supervised: Semantic Self-Consistency with External Feedback (P2 and P3)}
\label{sec:supervised-results}

\begin{table}[t]
\footnotesize
\caption{Comparison of supervised model understandability \underline{proxy P2} (Semantic Self-Consistency with LLM-as-a-Meta-Reviewer) against human actual binary understandability (ABU), and its agreement with P0 (LLM-as-a-Subject). Accuracy (A), Precision (P), Recall (R), F1, and AUC (\%). For (P2 vs P0), only F1 is shown.}
\vspace{-6pt}
\begin{tabular}{l|c|c|c|c|c||c}
\toprule
\multicolumn{1}{c|}{\textbf{$\langle$Model, Code$\rangle$ $\downarrow$}}
& \multicolumn{6}{c}{\textit{Evaluation Metrics} (\textit{in} \%)} \\ \cline{2-7}
\multicolumn{1}{c|}{}
& \multicolumn{5}{c||}{\textbf{P2 (LLM-as-a-Meta-Reviewer) vs ABU}}
& \textbf{P2 vs P0} \\ \cline{2-7}
\multicolumn{1}{c|}{}
& A & P & R & F1 & AUC & F1 \\ \midrule

Qwen2.5-14B         & 56.9 & 56.9 & 100.0 & 72.6 & 52.0 & 63.3 \\
WizardCoder-15B  & 56.9 & 56.9 & 100.0 & 72.6 & 53.7 & 65.2 \\
StarCoder2-15B    & 47.0 & 63.6 & 16.3  & 25.9 & 48.2 & 0.0  \\
CodeLlama-13B    & 50.3 & 62.8 & 31.4  & 41.9 & 50.0 & 10.5 \\ 
{\color{custom-blue}Qwen3-14B}    & 57.0 & 57.0 & 100.0 & 72.6 & 50.0 & 98.0 \\
{\color{custom-blue}Phi-4}    & 57.0 & 57.0 & 100.0 & 72.6 & 50.0 & 97.6 \\ 
\hline
GPT-3.5          & 58.3 & 58.2 & 95.3  & 72.2 & 54.1 & 94.0 \\
GPT-4            & 58.3 & 57.9 & 97.7  & \textbf{72.7} & 58.4 & 84.3 \\
{\color{custom-blue}GPT-4o} & 57.0 & 57.0 & 100.0 & 72.6 & 50.0 & 97.6 \\

\bottomrule
\end{tabular}
\label{tab:supervised}
\end{table}

As noted in Scalabrino {\em et al.}~\cite{scalabrino2021automatically}, the question-answering makes ABU a more reliable measure of human code understandability than PBU. Along similar lines, our supervised proxies (P2 and P3) use expert LLMs to provide external feedback, offering an automated procedure to intrinsically measure model-code understandability. Thus, we assess the alignment of P2 and P3 directly with~ABU. 

\subsubsection{Alignment of P2 (LLM-as-a-Meta-Reviewer) and ABU}

In Table~\ref{tab:supervised}, we report the results of comparing P2 with ABU. Using a global threshold of 0.45 and evaluating target model-generated summaries against a reference summary from an expert model ({\em i.e.}, with LLM-as-a-Meta-Reviewer), open-source models achieve an alignment with ABU in {\color{custom-blue}F1-scores ranging from 25.9\%--72.6\%. For closed-source models, the alignment is more consistent with F1-scores of 72.2--72.7\%}.

\input{figures/supervised-threshold-f1-auc}

Fig.~\ref{fig:supervised-threshold}({\em left}) further illustrates this performance across different thresholds. For smaller thresholds of similarity with the reference, 
the results of P2 closely resemble those of unsupervised semantic self-consistency (P1), as the effect of the reference constraint is negligible and the evaluation is primarily driven by self-consistency.
As the threshold increases, the evaluation against the reference summary becomes stricter. 
At the small peak across all models, cases of {\bf high-confidence hallucinations} captured by semantic self-consistency are filtered out, as they become dissimilar to the reference. Beyond this, however, increasing the threshold excludes the correctly aligned summaries, leading to a drop in F1-scores~(Fig.~\ref{fig:supervised-threshold}). 

{\color{custom-blue}
\underline{First}, in Table~\ref{tab:supervised} (Table~\ref{tab:qanda} and Table~\ref{tab:unsupervised} as well), the reported AUC values aim to capture~whether a proxy separates code snippets that humans judged understandable from those judged not understandable. As seen, they are close to 0.5. These {\bf AUC} and {\bf near-perfect recall} values indicate that for some model/proxy pairs, the {\em selected global threshold} places the predictor in a high-recall operating regime that classifies all programs understandable by humans as positive. This is especially~plausible since the ABU labels are skewed toward the positive class, while our evaluation casts alignment~as a binary classification. Thus, these cases should be interpreted as {\em threshold-sensitive~behavior}.

\underline{Second}, to provide further analysis on such weak-discrimination values, we present the changes~in AUC values over a range of BERTScore thresholds. As seen in Fig~\ref{fig:supervised-threshold} ({\em right}), changing the BERTScore threshold produces local peaks and troughs, but the curves remain close to 0.5 overall. This suggests that no single global threshold cleanly separates human-judged positive from negative cases. Instead, the proxy P2 scores for the two groups substantially overlap, so threshold adjustment mostly trades false positives against false negatives rather than producing robust discrimination.

\underline{Finally}, this motivates us to further stratify these AUC changes by different human reader subgroups. The subgroup-stratified results (Fig.~\ref{fig:p2-auc-by-role}) show that the aggregate weak discrimination in AUC values is caused by {\em mixing human reader groups}, and {\em the clearest positive separation in AUC values is obtained for the professional developers subgroup}. We will elaborate further in Section~\ref{sec:stratified}.
}

\subsubsection{Alignment of P3 (LLM-as-a-Judge) with ABU}



\begin{table}[t]
\footnotesize
\caption{Comparison of \underline{P3} (Semantic Self-Consistency with LLM-as-a-Judge) against 1) human Actual Binary Understandability (ABU) to assess {\color{custom-blue}{\bf behavioral alignment}}, and 2) P0 (LLM-as-a-Subject) to assess {\bf cross-proxy reliability}. We report Accuracy (A), Precision (P), Recall (R), F1, and AUC (\%) for P3 vs ABU. For P3 vs P0 in this version, only F1 is shown.}
\vspace{-9pt}
\begin{tabular}{l|c|c|c|c|c||c}
\toprule
\multicolumn{1}{c|}{\textbf{$\langle$Model, Code$\rangle$ $\downarrow$}}
& \multicolumn{6}{c}{\textit{Evaluation Metrics} (\textit{in} \%)} \\ \cline{2-7}
\multicolumn{1}{c|}{}
& \multicolumn{5}{c||}{\textbf{P3 (LLM-as-a-Judge) vs ABU}}
& \textbf{P3 vs P0} \\ \cline{2-7}
\multicolumn{1}{c|}{}
& A & P & R & F1 & AUC & F1 \\ \midrule

Qwen2.5-14B     & 56.3 & 58.1 & 83.7 & 68.6 & 51.9 & 59.8 \\
WizardCoder-15B & 48.3 & 65.4 & 19.8 & 30.4 & 53.0 & 20.2 \\
StarCoder2-15B   & 43.0 & 0.0  & 0.0  & 0.0  & 38.1 & 0.0  \\
CodeLlama-13B   & 43.0 & 0.0  & 0.0  & 0.0  & 38.1 & 0.0  \\ 
{\color{custom-blue}Qwen3-14B}   & 55.0 & 56.1 & 96.5 & 70.9 & 48.3 & 96.9 \\ 
{\color{custom-blue}Phi-4}   & 55.0 & 56.1 & 96.5 & 70.9 & 48.3 & 96.6 \\ 
\hline
GPT-3.5         & 57.6 & 57.4 & 98.8 & \textbf{72.6} & 55.0 & 92.9 \\
GPT-4           & 56.9 & 57.0 & 100.0 & \textbf{72.6} & 54.7 & \textbf{94.0} \\
{\color{custom-blue}GPT-4o} & 55.0 & 56.1 & 96.5 & 70.9 & 48.3 & 96.6 \\

\bottomrule
\end{tabular}
\label{tab:p3}
\end{table}

As seen in Table~\ref{tab:p3}, when using LLM-as-a-Judge for external feedback (P3), open-source models achieve substantially lower alignment with ABU, with {\color{custom-blue}F1-scores ranging from 0\% to 70.9\%}. This is notably weaker than their alignment under P2. A likely reason is that smaller open-source models tend to produce verbose intent summaries. In P2, such summaries can still be effectively evaluated against reference summaries from expert models using BERTScore, which is more tolerant to lexical variation.~However, in P3, the evaluation relies on an expert LLM directly judging whether a generated summary captures program intent. In this stricter setting, verbosity in summaries seems to cause noise that confuses the expert LLMs,~leading to lower alignment. The most effect is for StarCoder-15B and CodeLlama-13B, whose alignment drops to 0\% F1-score. In contrast, closed-source models, which produce concise intent summaries, maintain stable performance via P3 as well (72.6\% F1-score), consistent with their results under~P2.

\subsubsection{Reliability of Semantic Self-Consistency Proxy P3 for Model Code Understanding}
As seen in Table~\ref{tab:p3}, similar to P2, open-source models such as StarCoder2-15B and CodeLlama-13B have weak cross-proxy alignment between P3 and P0. {\color{custom-blue}In contrast, GPT-3.5, GPT-4, GPT-4o, Qwen3-14B, and Phi-4 show much higher alignment between P3 and P0.} 
Comparing Table~\ref{tab:supervised} and Table~\ref{tab:p3},
for GPT-3.5, P2 aligns more strongly with P0 than P3, suggesting that its code understandability is better reflected through reference-style supervision (where the expert LLM generates a reference summary). For GPT-4o, however: P3~aligns more strongly with P0 than P2, indicating that it is more reliable under direct supervision (where the expert LLM evaluates summaries against the program). These trends highlight that different proxies capture different aspects of model comprehension. Importantly, in Table~\ref{tab:unsupervised}, both P2 and P3 align well with the reliable proxy P0, thus they have potential applicability to new code benchmarks.


\begin{tcolorbox}[float=ht, top=2pt, bottom=2pt, left=5pt, right=5pt]
{\bf \em Findings (P2--P3):} 
\begin{enumerate}[topsep=0pt, itemsep=0pt, start=3, leftmargin=*]
    \item Incorporating external feedback through supervised proxies (P2 and P3) mitigates these limitations with P1 and yields more reliable estimates of model-code understandability.    
    \item Via both {\color{custom-blue}supervised behavioral proxies}, model understandability exhibits {\em moderate alignment} with human-code understandability.
    \item P0--P3 together offer a reliable and multi-dimensional lens for capturing model capabilities and assessing their alignment with general human readers' code understanding.
\end{enumerate}
\end{tcolorbox}

\section{Does Model Code Understandability Align with Human Reader Sub-Groups? (RQ2)}~\label{sec:stratified}

\begin{table}[ht]
\small
\caption{Stratified evaluation: variation of model code understandability based on human reader expertise.}
\vspace{-6pt}
\scalebox{0.81}{
\begin{tabular}{c|c|c|c|c||c|c|c|c||c|c|c|c||c|c|c|c}
\toprule
\multicolumn{1}{c|}{\multirow{3}{*}{\textbf{Model $\downarrow$}}} & \multicolumn{16}{c}{\textbf{Human Reader}}                                               \\ \cline{2-17} 
\multicolumn{1}{l|}{}          & \multicolumn{4}{c||}{\textit{Undergraduate}}            & \multicolumn{4}{c||}{\textit{Masters}}       & \multicolumn{4}{c||}{\textit{Ph.D.}}             & \multicolumn{4}{c}{\textit{\bf Professional}}          \\ \cline{2-17} 
\multicolumn{1}{c|}{\textbf{Proxy $\rightarrow$}}               & P0 & P1 & P2 & \multicolumn{1}{c||}{P3} & P0 & P1 & P2 & \multicolumn{1}{c||}{P3} & P0 & P1 & P2 & \multicolumn{1}{c||}{P3} & P0 & P1 & P2 & P3 \\ \midrule
Qwen2.5-14B         &   45.5      &   59.0     &      59.2    & \multicolumn{1}{c||}{59.0}     &  56.3     &  56.5      &   65.4   & \multicolumn{1}{c||}{58.3}   &  63.2  &    72.7     &   76.9   & \multicolumn{1}{c||}{60.9}     &    68.9    &   82.1   &   86.4   & 84.6       \\
WizardCoder-15B     &   39.9      &    60.6    &      59.2    & \multicolumn{1}{c||}{27.6}     &  54.5     &    66.7    &    65.4  & \multicolumn{1}{c||}{34.8}   &73.7   &   60.9      &   76.9   & \multicolumn{1}{c||}{28.6}     &   61.3     &   86.7   &    86.4  & 30.4       \\
StarCoder2-15B       &   7.7      &   52.9     &      21.4    & \multicolumn{1}{c||}{0}      &  19.0     &   57.1     &   36.4   & \multicolumn{1}{c||}{0}   & 16.7   &    76.9     &   15.4   & \multicolumn{1}{c||}{0}     &    23.3    &  84.7    &   26.7   & 0       \\
CodeLlama-13B       &   23.1      &   53.3     &      40.0    & \multicolumn{1}{c||}{0}     &  21.1     &   59.1     &    38.5  & \multicolumn{1}{c||}{0}   & 33.4   &   69.6      &   50.0   & \multicolumn{1}{c||}{0}     &    18.6    & 77.9     &   42.3   & 0       \\ 
{\color{custom-blue}Qwen3-14B}       &   58.0      &   60.0     &      59.2    & \multicolumn{1}{c||}{57.1}     &  66.7     &  65.4     & 65.4 & \multicolumn{1}{c||}{62.7}   & 72.0 & 76.9 & 76.9 & \multicolumn{1}{c||}{76.9}     & 86.0 & 85.1 & 86.4 & 85.1       \\
{\color{custom-blue}Phi-4}       &   58.0      &   59.7     &      59.2    & \multicolumn{1}{c||}{57.1}     &  66.7     &  64.0     & 65.4 & \multicolumn{1}{c||}{62.7}   & 72.0 & 66.7 & 76.9 & \multicolumn{1}{c||}{76.9}     & 86.0 & 85.7 & 86.4 & 85.1       \\
\hline
GPT-3.5             &   63.6      &   60.0     &      59.2    & \multicolumn{1}{c||}{60.0}     & 61.2      &   65.4     &  65.4    & \multicolumn{1}{c||}{65.4}   & 72.7   &   72.0      &  76.9    & \multicolumn{1}{c||}{72.0}     &   \cellcolor{gray!15}{86.7}     &   \cellcolor{gray!15}{85.1}   &  \cellcolor{gray!15}{86.4}    & \cellcolor{gray!15}{87.4}       \\
GPT-4               &    58.6     &   59.7     &     59.2     & \multicolumn{1}{c||}{59.2}     & 47.6     &   64.0     &   65.4   & \multicolumn{1}{c||}{65.4}   &  76.2  &   66.7      &    76.9  & \multicolumn{1}{c||}{77.0}     &   \cellcolor{gray!15}{77.3}     &   \cellcolor{gray!15}{85.7}   &    \cellcolor{gray!15}{86.4}  & \cellcolor{gray!15}{86.4}       \\
{\color{custom-blue}GPT-4o} & 60.0 & 58.0 & 59.2 & \multicolumn{1}{c||}{57.1} & 65.4 & 66.7 & 65.4 & \multicolumn{1}{c||}{62.7} & 72.0 & 72.0 & 76.9 & \multicolumn{1}{c||}{76.9} & \cellcolor{gray!15}{87.4} & \cellcolor{gray!15}{86.0} & \cellcolor{gray!15}{86.4} & \cellcolor{gray!15}{85.1} \\

\bottomrule
\end{tabular}
}
\label{tab:stratified}
\end{table}

In computing the proxies of human code understandability, Scalabrino {\em et al.}~\cite{scalabrino2021automatically} selected 63 Java developers of varying expertise, ranging from undergraduate students to professional developers. In this experiment, we analyze model code understandability, as captured by our proxies (P0--P3), by {\em partitioning each program's human reader ratings according to specific developer sub-groups}. 

Table~\ref{tab:stratified} presents the stratifying results. Across all proxies and models, we consistently observe that the alignment of model code understandability is {\bf strongest with professional developers}, followed by Ph.D. students, master's students, and then undergraduate students. For {\color{custom-blue}GPT-3.5, GPT-4 and GPT-4o}, the alignment of all proxies (P0--P3) with professional developers improves over the general population (as in Tables~\ref{tab:qanda}, ~\ref{tab:unsupervised}, and \ref{tab:supervised}) by {\color{custom-blue}an average of 14.1\%, 13.5\%, 14.3\%}, respectively. 

{\color{custom-blue}
For further analysis, we extend the threshold-based analysis via P2 (Section~\ref{sec:supervised-results} and Fig.~\ref{fig:supervised-threshold}) to the stratified setting. Fig.~\ref{fig:stratified-threshold} and Fig.~\ref{fig:p2-auc-by-role} show the changes in F1-score and AUC values over a range of thresholds for each subgroup of human readers, respectively. We make the following observations:

\underline{First}, as seen in Fig.~\ref{fig:stratified-threshold}, the F1-score results for all sub-groups are closer to those of the unsupervised proxy (P1) for smaller thresholds. Importantly, the professional group consistently maintains the strongest alignment across~thresholds as the trends in Fig.~\ref{fig:supervised-threshold} and Fig.~\ref{fig:stratified-threshold}  are similar.
}


\input{figures/stratified-supervised}


These findings are intuitive: since LLMs are primarily trained on code from open-source repositories ({\em e.g}, GitHub) and developer forums ({\em e.g.}, StackOverflow), their internalized representations of code semantics are more likely to reflect the perspectives of professional developers. As a result, the cognitive processes about code understanding in LLMs aligns most closely with that professional sub-group. A key implication of these findings, however, is that LLMs {\em cannot be naively applied across developer populations} without regard to their expertise (we expand on this in Section~\ref{sec:implications}).


{\color{custom-blue}
\underline{Second}, as seen in Fig.~\ref{fig:p2-auc-by-role}, the AUC results corresponding to the groups of undergraduate students and master students fluctuate near random discrimination, while those for the Ph.D. group often show unclear separation. Interestingly, {\em the AUC values corresponding to the professional-developer group show the clearest positive separation}. For this group, Qwen2.5-14B reaches an AUC value~of 0.706 at threshold of 0.61, StarCoder2-15B reaches 0.678 at 0.42, GPT-4 reaches 0.660 at 0.69, CodeLlama-13B reaches 0.634 at 0.42, Qwen3-14B reaches 0.612 at 0.66, WizardCoder-15B reaches 0.590 at 0.60, Phi-4 reaches 0.588 at 0.68, GPT-3.5 reaches 0.586 at 0.66, GPT-4o reaches 0.586 at 0.70. 

\underline{Third}, these AUC values are {\em consistent with the F1-scores in Fig.~\ref{fig:stratified-threshold}}, indicating that the behavioral alignment between model-code understandability (measured by P2) and human-code understandability (measured by ABU) is {\em closer for the subgroup of professional developers} than for the general reader population. This also gives {\em an explanation for the weak-discrimination AUC values} in Table~\ref{tab:supervised}. As aggregating the AUC values for the mixed human reader subgroups (undergraduate, master's, Ph.D. students, and professional developers) whose judgments on understandability vary greatly, we obtain a weak-discrimination AUC value, which does not indicate an uninformative~proxy.


\underline{Finally}, we also performed the same analyses for the behavioral alignment via other proxies (P0, P1, and P3) on the AUC changes over a range of thresholds for different human reader subgroups. Due to the space limit, we only present those corresponding figures on our project's website~\cite{code-understanding}. 

Despite the varied magnitudes across proxies and models, we made the same observations as in Fig.~\ref{fig:p2-auc-by-role} for P2: while the AUC curves for the subgroups of undergraduate, master's, and Ph.D. students remain near 0.5 or fluctuate around it, the AUC values corresponding to the professional developers exhibit clearer separation between positive and negative instances judged by humans. For example, for P1, the AUC value peaks at
0.743 for professional developers for Qwen2.5-14B, while that value for P3 reaches 0.61--0.66 for several models. This result also suggests that a separate threshold for each human reader subgroup is better than a global one over the mixed reader population.

}

\begin{tcolorbox}[top=2pt, bottom=2pt, left=5pt, right=5pt]
{\bf \em Findings:} 
\begin{enumerate}[topsep=0pt, itemsep=0pt, start=6, leftmargin=*]
    \item {\color{custom-blue}All BPMUs} P0--P3 point that model code understandability {\bf consistently and strongly aligns with professional developers'} while under-aligning with less experienced ones.
\end{enumerate}
\end{tcolorbox}

\section{Assessing the Effects of Role-Specific Conditioning (RQ3)}

\begin{table}[htpb]
\small
\caption{Assessing the effects of {\em role-specific prompting} in mimicking human reader expertise.}
\vspace{-6pt}
\scalebox{0.81}{
\begin{tabular}{c|c|c|c|c||c|c|c|c||c|c|c|c||c|c|c|c}
\toprule
\multicolumn{1}{c|}{\multirow{2}{*}{\textbf{Model $\downarrow$}}} & \multicolumn{16}{c}{\textbf{Human Reader}} \\ \cline{2-17} 
\multicolumn{1}{l|}{}                       & \multicolumn{4}{c||}{\textit{Undergraduate}}                                                                         & \multicolumn{4}{c||}{\textit{Masters}}                                                                               & \multicolumn{4}{c||}{\textit{Ph.D.}}                                                                                 & \multicolumn{4}{c}{\textit{Professional}}                                                      \\ \cline{2-17} 
\multicolumn{1}{c|}{\textbf{Proxy $\rightarrow$}}                       & {P0} & P1 & P2 & \multicolumn{1}{c||}{{P3}} & {P0} & P1 & P2 & \multicolumn{1}{c||}{{P3}} & {P0} & P1 & P2 & \multicolumn{1}{c||}{{P3}} & {P0} & P1 & P2 & {P3} \\ \midrule

Qwen2.5-14B                       & \multicolumn{1}{c|}{52.3}       & \multicolumn{1}{c|}{56.7}       & \multicolumn{1}{c|}{57.1}       & 62.2       & \multicolumn{1}{c|}{58.3}       & \multicolumn{1}{c|}{66.7}       & \multicolumn{1}{c|}{71.1}       & 61.2       & \multicolumn{1}{c|}{54.5}        & \multicolumn{1}{c|}{80.1}        & \multicolumn{1}{c|}{75.0}       & 66.7       & \multicolumn{1}{c|}{68.6}       & \multicolumn{1}{c|}{71.8}        & \multicolumn{1}{c|}{78.0}        & 74.6       \\ 
WizardCoder-15B                 & \multicolumn{1}{c|}{42.9}       & \multicolumn{1}{c|}{58.5}       & \multicolumn{1}{c|}{46.7}       & 21.2       & \multicolumn{1}{c|}{49.2}        & \multicolumn{1}{c|}{64.0}       & \multicolumn{1}{c|}{63.6}       & 28.6       & \multicolumn{1}{c|}{55.4}        & \multicolumn{1}{c|}{73.1}       & \multicolumn{1}{c|}{72.7}       & 0       & \multicolumn{1}{c|}{57.0}       & \multicolumn{1}{c|}{83.7}       & \multicolumn{1}{c|}{76.5}      & 14.3       \\ 
StarCoder2-15B                    & \multicolumn{1}{c|}{17.3}       & \multicolumn{1}{c|}{58.5}       & \multicolumn{1}{c|}{18.2}       & 0       & \multicolumn{1}{c|}{22.1}       & \multicolumn{1}{c|}{58.3}       & \multicolumn{1}{c|}{46.7}       & 0       & \multicolumn{1}{c|}{21.0}        & \multicolumn{1}{c|}{72.0}       & \multicolumn{1}{c|}{44.4}       & 0       & \multicolumn{1}{c|}{23.4}        & \multicolumn{1}{c|}{86.0}       & \multicolumn{1}{c|}{45.6}       & 0       \\ 
CodeLlama-13B                        & \multicolumn{1}{c|}{17.1}       & \multicolumn{1}{c|}{58.8}       & \multicolumn{1}{c|}{27.6}       & 0       & \multicolumn{1}{c|}{15.2}       & \multicolumn{1}{c|}{59.6}       & \multicolumn{1}{c|}{46.2}       & 0       & \multicolumn{1}{c|}{24.3}       & \multicolumn{1}{c|}{76.8}       & \multicolumn{1}{c|}{53.3}       & 0       & \multicolumn{1}{c|}{24.3}       & \multicolumn{1}{c|}{82.4}       & \multicolumn{1}{c|}{33.3}       & 0       \\
{\color{custom-blue}Qwen3-14B}       & \multicolumn{1}{c|}{58.0}       & \multicolumn{1}{c|}{60.0}       & \multicolumn{1}{c|}{59.2}       & 57.1       & \multicolumn{1}{c|}{66.7}       & \multicolumn{1}{c|}{66.7}       & \multicolumn{1}{c|}{65.4}       & 62.7       & \multicolumn{1}{c|}{72.0}       & \multicolumn{1}{c|}{76.9}       & \multicolumn{1}{c|}{76.9}       & 76.9       & \multicolumn{1}{c|}{86.0}       & \multicolumn{1}{c|}{85.1}       & \multicolumn{1}{c|}{86.4}       & 85.1       \\
{\color{custom-blue}Phi-4}       & \multicolumn{1}{c|}{58.8}       & \multicolumn{1}{c|}{60.0}       & \multicolumn{1}{c|}{59.2}       & 59.2       & \multicolumn{1}{c|}{66.7}       & \multicolumn{1}{c|}{66.7}       & \multicolumn{1}{c|}{65.4}       & 62.7       & \multicolumn{1}{c|}{72.0}       & \multicolumn{1}{c|}{76.9}       & \multicolumn{1}{c|}{76.9}       & 76.9       & \multicolumn{1}{c|}{86.0}       & \multicolumn{1}{c|}{85.1}       & \multicolumn{1}{c|}{86.4}       & 86.4       \\ \hline
GPT-3.5                         & \multicolumn{1}{c|}{69.7}       & \multicolumn{1}{c|}{55.9}       & \multicolumn{1}{c|}{55.4}       & 60.0       & \multicolumn{1}{c|}{68.2}       & \multicolumn{1}{c|}{61.2}       & \multicolumn{1}{c|}{62.5}       & 65.4       & \multicolumn{1}{c|}{70.0}        & \multicolumn{1}{c|}{76.8}        & \multicolumn{1}{c|}{72.0}       & 72.0       & \multicolumn{1}{c|}{71.9}        & \multicolumn{1}{c|}{86.0}       & \multicolumn{1}{c|}{81.9}       & 87.4       \\ 
GPT-4                               & \multicolumn{1}{c|}{65.2}       & \multicolumn{1}{c|}{53.7}       & \multicolumn{1}{c|}{56.3}       & 59.2       & \multicolumn{1}{c|}{59.1}       & \multicolumn{1}{c|}{61.2}       & \multicolumn{1}{c|}{63.8}       & 65.4       & \multicolumn{1}{c|}{64.6}        & \multicolumn{1}{c|}{74.2}       & \multicolumn{1}{c|}{69.6}       & 76.9       & \multicolumn{1}{c|}{71.8}        & \multicolumn{1}{c|}{79.0}       & \multicolumn{1}{c|}{82.4}       & 86.4       \\ 
{\color{custom-blue}GPT-4o} & \multicolumn{1}{c|}{58.8} & \multicolumn{1}{c|}{58.0} & \multicolumn{1}{c|}{59.2} & 58.0 & \multicolumn{1}{c|}{66.7} & \multicolumn{1}{c|}{65.4} & \multicolumn{1}{c|}{65.4} & 65.4 & \multicolumn{1}{c|}{72.0} & \multicolumn{1}{c|}{76.9} & \multicolumn{1}{c|}{76.9} & 76.9 & \multicolumn{1}{c|}{86.0} & \multicolumn{1}{c|}{86.4} & \multicolumn{1}{c|}{86.4} & 87.4 \\

\bottomrule
\end{tabular}
}
\label{tab:role}
\end{table}

As noted in Section~\ref{sec:stratified}, in the absence of any role designation, model code understandability aligns most with the code understandability of professional developers among the human reader sub-populations. In this experiment, we investigate if {\em role-conditioned prompting} can help capture the perspectives of these sub-populations. This idea follows from previous work, which used phrases like {\em ``Assume I am a novice programmer''}~\cite{DBLP:journals/corr/abs-2409-14368} or {\em ``I want you to act as a developer''}~\cite{DBLP:journals/tosem/DongJJL24} to steer the LLMs into adopting specific persona and, in turn, adjust the tone or complexity of their outputs. To enable such role designations, we prefixed each prompt with an explicit persona: "{\em You are a bachelor's student /master's student / Ph.D. student / senior software engineer ...}$\langle${\code{role-description}}$\rangle$".


In Table~\ref{tab:role}, we present the results for the alignment of proxies P0--P3 with ABU for each human reader sub-population under explicit role prompts. Compared to baselines without any role designations (Table~\ref{tab:stratified}), the results show that {\em all proxies remain largely unchanged across all human reader categories}. This indicates that models are {\em not sensitive to role-specific conditioning}, even performing worse in some cases under these settings; and such prompts {\em do not reliably make the model's understanding align with the target group's understanding}. 
These findings suggest that additional mechanisms, such as few-shot learning or targeted fine-tuning are required for LLMs to reliably adopt such personas. A direct implication of these findings lies in educational or pair-programming  contexts (further expanded in Section~\ref{sec:implications}), where relying on role prompts alone may not meaningfully tailor model behavior to diverse learner backgrounds.


\begin{tcolorbox}[top=2pt, bottom=2pt, left=5pt, right=5pt]
{\bf \em Findings:} 
\begin{enumerate}[topsep=0pt, itemsep=0pt, start=7, leftmargin=*]
    \item Role-conditioned prompting does not help large language models reliably mimic the perspectives of developers with varying expertise.
\end{enumerate}
\end{tcolorbox}




\section{Implications}
\label{sec:implications}

\subsection{Software Engineering Research}


{\em Theorizing relational code understandability.} A key implication of our findings is the need to reconceptualize code quality attributes such as understandability through a {\bf relational lens}, where quality is not solely a property of the code but emerges from its interaction with the {\bf reader}, which could be {\em humans or models}. Existing approaches have emphasized on two directions: {\bf human-centric} (where models need to be aligned with humans) and {\bf code-centric} (understandability depends only on code properties, {\em e.g.}, cyclomatic complexity without considering readers). Our results challenge this view, showing that understandability varies greatly across professional developers, students, and LLMs, and that intrinsic metrics fail to capture these differences. Future research should develop new frameworks that explicitly model understandability as relational, accounting for (i) {\bf reader expertise} (novice vs. expert), (ii) {\bf reader type} (human, LLM, or program analysis tool), and (iii) {\bf contextual factors} (e.g., task, prior knowledge, prompt conditioning). This shift opens the door to relational quality metrics, such as measuring how well code is understood by both humans and LLMs, or quantifying disparities across reader groups.

{\em Hybrid human-AI code comprehension models for code review}. 
First, our findings suggest that LLMs align more closely with professional developers than beginners, indicating they are {\em collaborators for expert programmers rather than replacements across all levels of human comprehension}.
This motivates research into building frameworks for hybrid {\bf human--AI program comprehension}, where humans and LLMs work collaboratively, complementing each other's strengths.
For instance, in {\em code review}, LLMs can flag readability or maintainability issues, while humans can validate them against project-specific conventions. In {\em documentation and summarization}, LLMs can generate initial descriptions, while humans can refine them for clarity. 
Second, beyond task-level collaboration, hybrid models could support {\bf interactive workflows} where humans query an LLM and the model adapts its explanations to the user's expertise and feedback. This addresses the limitation that role-conditioned prompting alone fails to replicate novice or expert perspectives.

{\em Human-Computer Interaction}. A promising research direction is the design of {\bf collaborative frameworks, protocols, and interfaces} that facilitate shared responsibility between humans and models. Ultimately, the question shifts from whether LLMs can replace humans to how {\bf multi-reader systems} achieve higher-quality outcomes. 
Moreover, interfaces for code review or {\bf pair programming} could embed as an additional participant that flexibly adapts its explanations depending on which collaborator it is addressing.

{\em Multi-reader evaluation methodologies and benchmarks}. 
The current one-dimensional evaluation overlooks that code is interpreted differently by diverse readers. Future work should develop {\bf multi-reader evaluation methodologies} to capture and compare these perspectives, quantifying alignments and divergences across reader groups. This shift requires new {\bf benchmarks and datasets} built with reader diversity in mind. Instead of only general tasks or snippets, datasets should include comprehension challenges faced by novices and contexts where LLMs or automated tools struggle. Examples of such benchmark tasks for multi-reader evaluation include:


$\bullet$ {\em Mixed human-LLM code review corpora}: Curated repositories of real code reviews augmented with LLM-generated feedback on the same commits. This corpora would also reveal blind spots and biases in either human or model feedback.

$\bullet$ {\em Multi-dimensional comprehension tasks}: Benchmarks where participants and models must evaluate code not only on functional correctness but also understandability. This encourages~models to approximate the broader notion of quality beyond correctness, in line with human judgments.

{\em Empirical studies with scalable LLM proxies}.
Our results indicate the potential use of LLMs as scalable proxies in empirical studies of code understandability. Since their assessments align more with professional developers than novices, LLMs could serve as cost-effective stand-ins for expert judgments in large-scale experiments, where recruiting professionals is often a bottleneck. For instance, they could provide preliminary ratings to filter examples or generate large annotated datasets before investing in smaller, high-quality human studies. Still, LLMs cannot fully replace human participants. Human judgments remain indispensable for novice comprehension, domain-specific reasoning, and evaluating educational effectiveness. Future work should therefore explore hybrid study designs that combine the scalability of LLM evaluations with the depth of human studies, leveraging efficiency without losing nuanced insight that only humans can provide.

\subsection{Developers and Practitioners, and Training}
\label{sec:dev}


For professional developers, the high alignment between LLM assessments and expert judgments suggests that models can serve as {\bf reliable collaborators} in comprehension-heavy tasks such as code review and documentation, reducing cognitive effort by surfacing concerns and generating accurate documentation.
However, weaker alignment for novices and non-professionals shows that LLMs are biased toward expert comprehension, often missing beginner challenges.
Thus, caution is needed when applying LLMs for {\bf mentorship, onboarding, or training}, as model explanations may confuse for new learners. Organizations might integrate LLMs into {\bf onboarding pipelines} in a controlled way, e.g., scaffolding expert-level documentation while human mentors provide novice-friendly explanations. In teams, professionals can adopt {\bf dual-perspective workflows}, combining LLM feedback for expert-level comprehension with mentoring practices to bridge novice gaps. Finally, professionals could use LLMs as {\bf diagnostic mirrors}, noting where model alignment indicates clarity for experts and misalignment signals potential novice difficulties.

\subsection{Software Engineering Education}
\label{sec:edu}


The alignment gap for students shows that current LLMs can not directly replace human mentorship in SE education as their judgments reflect expert reasoning. \underline{First}, this underscores the importance of {\bf human instructors in scaffolding learning}, especially in addressing misconceptions LLMs may miss. \underline{Second}, it points to {\bf hybrid teaching aids}, where LLM feedback 
is integrated into classrooms but moderated by teachers to contextualize outputs and reframe them for students' level. \underline{Third}, educators could also design {\bf interventions that treat model limitations as learning opportunities}, using mismatches between LLM and novice reasoning to foster critical thinking~and~reflection, and to evaluate the reliability of AI-provided feedback. \underline{Finally}, there is potential for {\bf customized learning}: fine-tuned 
models might provide feedback tailored to skill levels. ``Dual-mode teaching assistants'' could pair expert-oriented explanations with simplified reasoning for beginners, allowing students to transition gradually from novice-friendly to expert-level perspectives.

\section{Threats to Validity}
\label{sec:threats}

\paragraph{External Validity} 
{\color{custom-blue}
The main goal of our study is to assess the behavioral alignment between humans and models.
Thus, our evaluation requires a dataset with {\em human-grounded understandability labels}, ideally with~correctness-based verification questions across reader groups. Such datasets are currently scarce. For this reason, we build on Scalabrino {\em et al.}'s data~\cite{scalabrino2021automatically}. This constraint could be a threat to the generality of our conclusion, as we used the ground truth labels and 
the behavioral proxies (e.g., PBU and ABU) in that benchmark. 
Moreover, it may not fully represent the diversity of real-world programming tasks or developer populations. The participant pool consisted of students and professional developers, which may not capture the full spectrum of developer expertise.
}

Likewise, we evaluated on a selected set of target LLMs, and with other models/architectures, may exhibit different behaviors. These factors should be considered when generalizing our findings, though the consistent trends across proxies and models provide confidence in generalization.

\vspace{-7pt}
\paragraph{Internal Validity} A common threat in empirical studies with LLMs is potential data leakage. In our case, however, the human judgments used for correctness-based evaluations in P0 are unlikely to appear in pretraining data. Furthermore, the alignment of P1--P3 with P0 reinforces that semantic self-consistency reflects genuine comprehension signals rather than surface-level memorization (Section~\ref{sec:prelim}). Another possible source of bias arises from prompt design and experimental setup; to mitigate this, we applied consistent prompting strategies across all models.



\vspace{-7pt}
\paragraph{Construct Validity} 
Based on the operationalization of human and model code understandability: (1) {\color{custom-blue}for human judgments, behavioral proxies of code understandability such as PBU and ABU 
used in our study capture distinct but limited facets of comprehension, which may not fully reflect the multi-dimensional nature of human code understanding such as cognitive or neural mechanisms}; (2) for models, semantic self-consistency and Q\&A correctness similarly approximate, but are not perfect. 
Moreover, we use three state-of-the-art LLMs as expert models, and thus, did not consider them as target models. However, we included multiple open and closed-source models in the study.

\section{Related Work}

\paragraph{Human-Centric Code Understandability}
Brooks' seminal work ``Towards a theory of the~comprehension of computer programs'' framed it as building mental models using knowledge structures (\textit{beacons}) shaped by prior experience~\cite{BROOKS1983543}. Rajlich and Wilde later stressed the role of high-level \textit{concepts}, arguing that developers map code to domain concepts to drive understanding~\cite{rajlich2002role}. These studies highlight that comprehension depends on individual knowledge and strategy, a view reinforced by recent evidence of substantial \emph{inter-personal variation} in understandability judgments~\cite{scalabrino2021automatically}.

Numerous metrics have been proposed to measure code understandability (or related properties like complexity and readability). Complexity metrics (McCabe's cyclomatic complexity~\cite{mccabe1976complexity}, Halstead's volume~\cite{halstead1977elements}, Cognitive Complexity~\cite{cognitivecomplexity2017}) were assumed to reflect ease of understanding, but other studies refute this. Scalabrino \textit{et al.}~\cite{scalabrino2021automatically} tested 121 metrics and found none significantly correlated with perceived or actual understandability. Trockman \textit{et al.}~\cite{Trockman2018} combined metrics with statistical modeling, yielding modest predictive power. Another line of work used \textit{readability} as a proxy, with Buse and Weimer predicting human judgments from surface features~\cite{Buse2010}, but readability captures only stylistic clarity, not true comprehension. These studies show that understandability emerges from reader-dependent interactions beyond what static metrics capture.

\vspace{-6pt}
\paragraph{Individual Differences: Novice vs. Expert Understanding} 
Research in SE and computing education show that a reader's expertise strongly shapes code comprehension. Eye-tracking studies by Crosby and Stelovsky found novices read code linearly, like texts, while experts use non-linear strategies, scanning and focusing on key sections~\cite{Crosby1990}. Busjahn \textit{et al.}~\cite{Busjahn2015} further showed experts exhibit more non-linear gaze patterns than novices' sequential reading. These studies support our relational view: what is ``understandable'' code differs greatly between experts and novices.


\vspace{-4pt}
\paragraph{Machine-Based Code Understanding and Summarization} 
Early work applied text summarization to source code, generating short descriptions from code as input~\cite{haiduc2010summarizing,rodeghero2014eyetracking}. LLMs now achieve strong results in code summarization~\cite{Iyer2016,ahmad2020ast,8811932,Feng2020,Wang2021CodeT5,Lu2021CodexGlue}, though they are trained to predict comments or ensure correctness rather than match \emph{cognitive} understanding. Still, their performance suggests they capture substantial semantics and can serve as approximate ML/LLM-based readers of code. In parallel, large general-purpose LLMs show impressive code understanding and generation abilities~\cite{Brown2020,Chen2021,Wang2021CodeT5}. It remains unclear how closely their ``understanding'' aligns with humans, since they optimize for proxies such as output or stylistic patterns that may not match human intuitions.

\vspace{-4pt}
\paragraph{Model Explainability/Understanding} 
Previous XAI research has focused on explaining how large language models understand code, {\em e.g.}, syntactic and semantic properties~\cite{DBLP:conf/kbse/LopezWCS22,DBLP:conf/blackboxnlp/TroshinC22,DBLP:conf/icse/NorthAB25}. In contrast, we propose semantic self-consistency to quantify the understandability of code in LLMs.

Wang \textit{et al.}~\cite{Wang2023self} showed that prompting an LLM to generate multiple reasoning chains and selecting the most consistent answer improves accuracy on complex Q\&A tasks. The intuition is that true ``understanding'' yields consistent answers across reasoning paths, while inconsistency suggests a shallow grasp. We draw on this in our first proxy and extend it with \textit{cross-model agreement}: comparing a model's summaries to those from strong ``expert'' models (e.g., GPT-4, Claude) to gauge whether weaker models capture the same key behavior and intent. 

Recent work has advanced self-consistency and conformance methods. Wang \textit{et al.}~\cite{wang2023selfconsistency} introduced \textit{self-consistency}, sampling diverse reasoning paths and choosing the most frequent answer to boost accuracy. Chen \textit{et al.}~\cite{chen2023usc} generalized with \textit{Universal Self-Consistency} (USC) for open-ended tasks. Taubenfeld \textit{et al.}~\cite{taubenfeld2025cisc} improved efficiency by weighting paths in confidence. Prasad \textit{et al.}~\cite{prasad2024scpo} proposed \textit{Self-Consistency Preference Optimization} (ScPO) to directly optimize for consistency. Reward models are stabilized via \textit{Self-Consistent Internal Rewards} (SCIR)~\cite{zhou2025scir}. Other directions include dynamic sampling guided by reasoning paths~\cite{wan2025reasoningawareselfconsistencyleveraging}, adaptive sampling with~cross-verification, and including human-like cognitive behaviors such as double-checking and heuristic relaxation.

\section{Conclusion}
\label{sec:conclusion}

Existing approaches have emphasized on two directions: {\bf human-centric} (where models need to be aligned with humans) and {\bf code-centric} (understandability depends only on code properties, {\em e.g.}, cyclomatic complexity without considering readers).

In this work, we re-conceptualized code understandability as a relational property between readers and programs, extending the notion of ``reader'' to include both humans and large language models (LLMs). To operationalize this perspective, we introduced {\color{custom-blue}multiple behavioral proxies for model code understandability (BPMU), examining whether these proxies align with the proxies for human code understandability}. Overall, we found that LLMs align more closely with professional developers than undergraduate/graduate
students. We demonstrate that (supervised) semantic self-consistency provides a reliable and extensible framework for quantifying model code understandability. These findings underscore the need to treat LLMs as collaborators rather than replacements in program comprehension, exploring interactive workflows and multi-reader systems where humans and models complementarily contribute.

\section{Data Availability}
Our data and code are available at ~\cite{code-understanding}.



\balance

\bibliographystyle{ACM-Reference-Format}

\bibliography{references}

@MISC{ChatGPT,
        TITLE     = {{OpenAI}},
        HOWPUBLISHED = {https://openai.com/},
        KEY       = {ChatGPT},
}

@misc{code_llama,
      title={Code Llama: Open Foundation Models for Code}, 
      author={Baptiste Roziere and Jonas Gehring and Fabian Gloeckle and Sten Sootla and Itai Gat and Xiaoqing Ellen Tan and Yossi Adi and Jingyu Liu and Tal Remez and Jeremy Rapin and Artyom Kozhevnikov and Ivan Evtimov and Joanna Bitton and Manish Bhatt and Cristian Canton Ferrer and Aaron Grattafiori and Wenhan Xiong and Alexandre Defossez and Jade Copet and Faisal Azhar and Hugo Touvron and Louis Martin and Nicolas Usunier and Thomas Scialom and Gabriel Synnaeve},
      year={2023},
      eprint={2308.12950},
      archivePrefix={arXiv},
      primaryClass={cs.CL}
}

@ARTICLE{scalabrino2021automatically,
  author={Scalabrino, Simone and Bavota, Gabriele and Vendome, Christopher and Linares-Vasquez, Mario and Poshyvanyk, Denys and Oliveto, Rocco},
  journal={IEEE Transactions on Software Engineering}, 
  title={Automatically Assessing Code Understandability}, 
  year={2021},
  volume={47},
  number={3},
  pages={595-613},
  doi={10.1109/TSE.2019.2901468}}

@inproceedings{rajlich2002role,
author = {Rajlich, V\'{a}clav and Wilde, Norman},
title = {The Role of Concepts in Program Comprehension},
year = {2002},
isbn = {0769514952},
publisher = {IEEE Computer Society},
address = {USA},
booktitle = {Proceedings of the 10th International Workshop on Program Comprehension},
pages = {271},
series = {IWPC '02}
}

@article{BROOKS1983543,
title = {Towards a theory of the comprehension of computer programs},
journal = {International Journal of Man-Machine Studies},
volume = {18},
number = {6},
pages = {543-554},
year = {1983},
issn = {0020-7373},
doi = {https://doi.org/10.1016/S0020-7373(83)80031-5},
url = {https://www.sciencedirect.com/science/article/pii/S0020737383800315},
author = {Ruven Brooks}
}

@inproceedings{zhang2019bertscore,
  title     = {BERTScore: Evaluating Text Generation with BERT},
  author    = {Zhang, Tianyi and Kishore, Varsha and Wu, Felix and Weinberger, Kilian Q. and Artzi, Yoav},
  booktitle = {International Conference on Learning Representations (ICLR)},
  year      = {2020},
  url       = {https://openreview.net/forum?id=SkeHuCVFDr}
}

@misc{qwen2025qwen25technicalreport,
      title={Qwen2.5 Technical Report}, 
      author={Qwen and : and An Yang and Baosong Yang and Beichen Zhang and Binyuan Hui and Bo Zheng and Bowen Yu and Chengyuan Li and Dayiheng Liu and Fei Huang and Haoran Wei and Huan Lin and Jian Yang and Jianhong Tu and Jianwei Zhang and Jianxin Yang and Jiaxi Yang and Jingren Zhou and Junyang Lin and Kai Dang and Keming Lu and Keqin Bao and Kexin Yang and Le Yu and Mei Li and Mingfeng Xue and Pei Zhang and Qin Zhu and Rui Men and Runji Lin and Tianhao Li and Tianyi Tang and Tingyu Xia and Xingzhang Ren and Xuancheng Ren and Yang Fan and Yang Su and Yichang Zhang and Yu Wan and Yuqiong Liu and Zeyu Cui and Zhenru Zhang and Zihan Qiu},
      year={2025},
      eprint={2412.15115},
      archivePrefix={arXiv},
      primaryClass={cs.CL},
      url={https://arxiv.org/abs/2412.15115}, 
}

@misc{luo2025wizardcoderempoweringcodelarge,
      title={WizardCoder: Empowering Code Large Language Models with Evol-Instruct}, 
      author={Ziyang Luo and Can Xu and Pu Zhao and Qingfeng Sun and Xiubo Geng and Wenxiang Hu and Chongyang Tao and Jing Ma and Qingwei Lin and Daxin Jiang},
      year={2025},
      eprint={2306.08568},
      archivePrefix={arXiv},
      primaryClass={cs.CL},
      url={https://arxiv.org/abs/2306.08568}, 
}

@misc{lozhkov2024starcoder2stackv2,
      title={StarCoder 2 and The Stack v2: The Next Generation}, 
      author={Anton Lozhkov and Raymond Li and Loubna Ben Allal and Federico Cassano and Joel Lamy-Poirier and Nouamane Tazi and Ao Tang and Dmytro Pykhtar and Jiawei Liu and Yuxiang Wei and Tianyang Liu and Max Tian and Denis Kocetkov and Arthur Zucker and Younes Belkada and Zijian Wang and Qian Liu and Dmitry Abulkhanov and Indraneil Paul and Zhuang Li and Wen-Ding Li and Megan Risdal and Jia Li and Jian Zhu and Terry Yue Zhuo and Evgenii Zheltonozhskii and Nii Osae Osae Dade and Wenhao Yu and Lucas Krauß and Naman Jain and Yixuan Su and Xuanli He and Manan Dey and Edoardo Abati and Yekun Chai and Niklas Muennighoff and Xiangru Tang and Muhtasham Oblokulov and Christopher Akiki and Marc Marone and Chenghao Mou and Mayank Mishra and Alex Gu and Binyuan Hui and Tri Dao and Armel Zebaze and Olivier Dehaene and Nicolas Patry and Canwen Xu and Julian McAuley and Han Hu and Torsten Scholak and Sebastien Paquet and Jennifer Robinson and Carolyn Jane Anderson and Nicolas Chapados and Mostofa Patwary and Nima Tajbakhsh and Yacine Jernite and Carlos Muñoz Ferrandis and Lingming Zhang and Sean Hughes and Thomas Wolf and Arjun Guha and Leandro von Werra and Harm de Vries},
      year={2024},
      eprint={2402.19173},
      archivePrefix={arXiv},
      primaryClass={cs.SE},
      url={https://arxiv.org/abs/2402.19173}, 
}

@misc{gemini25_2025,
  title        = {Gemini 2.5: Our most intelligent AI model},
  author       = {{Google DeepMind}},
  howpublished = {\url{https://blog.google/technology/google-deepmind/gemini-model-thinking-updates-march-2025/}},
  year         = {2025},
  note         = {Accessed 2025-09-10}
}

@misc{claude_opus41_2025,
  title        = {Claude Opus 4.1},
  author       = {{Anthropic}},
  howpublished = {\url{https://www.anthropic.com/claude/opus}},
  year         = {2025},
  note         = {Accessed 2025-09-10}
}

@article{Buse2010,
author = {Raymond P. L. Buse and Westley R. Weimer},
title = {Learning a Metric for Code Readability},
journal = {IEEE Trans. Software Eng.},
volume = {36},
number = {4},
pages = {546--558},
year = {2010},
doi = {10.1109/TSE.2009.70}
}

@inproceedings{Trockman2018,
author = {Asher Trockman and Keenen Cates and Mark Mozina and Tuan Nguyen and Christian Kastner and Bogdan Vasilescu},
title = {{\textquotedblleft}Automatically Assessing Code Understandability{\textquotedblright} Reanalyzed: Combined Metrics Matter},
booktitle = {Proc. 15th Int. Conf. Mining Software Repositories (MSR)},
pages = {46--57},
year = {2018},
publisher = {ACM}
}

@article{Crosby1990,
author = {Martha E. Crosby and Jan Stelovsky},
title = {How do we read algorithms? A case study},
journal = {{IEEE} Computer},
volume = {23},
number = {1},
pages = {25--35},
year = {1990}
}

@inproceedings{Busjahn2015,
author = {Teresa Busjahn and Roman Bednarik and Andrew Begel and Martha Crosby and James Paterson and Carsten Schulte and Bonita Sharif and Sascha Tamm},
title = {Eye Movements in Code Reading: Relaxing the Linear Order},
booktitle = {Proc. 23rd {IEEE} Int. Conf. Program Comprehension (ICPC)},
pages = {255--265},
year = {2015},
publisher = {IEEE},
doi = {10.1109/ICPC.2015.36}
}

@inproceedings{Iyer2016,
author = {Srinivasan Iyer and Ioannis Konstas and Alvin Cheung and Luke Zettlemoyer},
title = {Summarizing Source Code using a Neural Attention Model},
booktitle = {Proc. 54th Annual Meeting of the Association for Computational Linguistics (ACL)},
pages = {2073--2083},
year = {2016},
publisher = {Association for Computational Linguistics}
}

@inproceedings{Feng2020,
author = {Zhangyin Feng and Daya Guo and Duyu Tang and Nan Duan and Xiaocheng Feng and Ming Gong and Linjun Shou and Bing Qin and Ting Liu and Daxin Jiang and Ming Zhou},
title = {CodeBERT: A Pre-Trained Model for Programming and Natural Languages},
booktitle = {Findings of the Association for Computational Linguistics: {EMNLP} 2020},
pages = {1536--1547},
year = {2020},
publisher = {Association for Computational Linguistics}
}

@inproceedings{Wang2021CodeT5,
author = {Yue Wang and Weishi Wang and Shafiq Joty and Steven C.~H. Hoi},
title = {{CodeT5}: Identifier-Aware Unified Pre-Trained Encoder-Decoder Models for Code Understanding and Generation},
booktitle = {Proc. 2021 Conf. Empirical Methods in Natural Language Processing (EMNLP)},
pages = {8696--8710},
year = {2021},
publisher = {Association for Computational Linguistics}
}

@inproceedings{Brown2020,
author = {Tom B. Brown and Benjamin Mann and Nick Ryder and Melanie Subbiah and Jared Kaplan and Prafulla Dhariwal and Arvind Neelakantan and Pranav Shyam and Girish Sastry and Amanda Askell and et al.},
title = {Language Models are Few-Shot Learners},
booktitle = {Advances in Neural Information Processing Systems (NeurIPS) 33},
pages = {1877--1901},
year = {2020}
}

@article{Chen2021,
author = {Mark Chen and Jerry Tworek and Heewoo Jun and Qiming Yuan and Henrique Ponde de Oliveira Pinto and Jared Kaplan and Harri Edwards and Yuri Burda and et al.},
title = {Evaluating Large Language Models Trained on Code},
journal = {arXiv preprint arXiv:2107.03374},
year = {2021}
}

@inproceedings{Wang2023self,
author = {Xuezhi Wang and Jason Wei and Dale Schuurmans and Quoc V. Le and Ed H. Chi and Sharan Narang and Aakanksha Chowdhery and Denny Zhou},
title = {Self-Consistency Improves Chain-of-Thought Reasoning in Language Models},
booktitle = {Proc. Int. Conf. Learning Representations (ICLR)},
year = {2023}
}

@inproceedings{haiduc2010summarizing,
  title={On the Use of Automated Text Summarization Techniques for Summarizing Source Code},
  author={Haiduc, Sonia and Aponte, Jairo and Moreno, Laura and Marcus, Andrian},
  booktitle={Proceedings of the 17th Working Conference on Reverse Engineering (WCRE)},
  year={2010},
  pages={35--44},
  address={Beverly, MA, USA},
  month={October},
  publisher={IEEE Computer Society},
  doi={10.1109/WCRE.2010.13}
}

@inproceedings{rodeghero2014eyetracking,
author = {Rodeghero, Paige and McMillan, Collin and McBurney, Paul W. and Bosch, Nigel and D'Mello, Sidney},
title = {Improving automated source code summarization via an eye-tracking study of programmers},
year = {2014},
isbn = {9781450327565},
publisher = {Association for Computing Machinery},
address = {New York, NY, USA},
url = {https://doi.org/10.1145/2568225.2568247},
doi = {10.1145/2568225.2568247},
booktitle = {Proceedings of the 36th International Conference on Software Engineering},
pages = {390-401},
numpages = {12},
location = {Hyderabad, India},
series = {ICSE 2014}
}

@inproceedings{ahmad2020ast,
  title={A Transformer-based Approach for Source Code Summarization},
  author={Ahmad, Wasi Uddin and Chakraborty, Saikat and Ray, Baishakhi and Chang, Kai-Wei},
  booktitle={Proceedings of the 58th Annual Meeting of the Association for Computational Linguistics (ACL)},
  year={2020},
  pages={4998--5007},
  address={Online},
  publisher={Association for Computational Linguistics},
  doi={10.18653/v1/2020.acl-main.449}
}

@inproceedings{lu2021codexglue,
  title={CodeXGLUE: A Machine Learning Benchmark Dataset for Code Understanding and Generation},
  author={Lu, Shuai and Guo, Daya and Ren, Shuo and Huang, Junjie and Svyatkovskiy, Alexey and Fu, Shengyu and Li, M. and Zhou, L. and Tufano, M. and Drain, D. and Deng, S. and Clement, C. and Sundaresan, N. and Deng, G. and Fu, W. and Liu, A. and Liu, D. and Tang, Y. and Zhang, J. and Chandra, S. and Zhou, M. and Gong, M.},
  booktitle={Proceedings of the 35th Conference on Neural Information Processing Systems (NeurIPS), Datasets and Benchmarks Track},
  year={2021},
  url={https://arxiv.org/abs/2102.04664}
}

@article{mccabe1976complexity,
  title={A Complexity Measure},
  author={McCabe, Thomas J.},
  journal={IEEE Transactions on Software Engineering},
  volume={SE-2},
  number={4},
  pages={308--320},
  year={1976},
  publisher={IEEE}
}

@book{halstead1977elements,
  title={Elements of Software Science (Operating and Programming Systems Series)},
  author={Halstead, Maurice H.},
  year={1977},
  publisher={Elsevier Science Inc.}
}

@inproceedings{cognitivecomplexity2017,
author = {Campbell, G. Ann},
title = {Cognitive complexity: an overview and evaluation},
year = {2018},
isbn = {9781450357135},
publisher = {Association for Computing Machinery},
address = {New York, NY, USA},
url = {https://doi.org/10.1145/3194164.3194186},
doi = {10.1145/3194164.3194186},
booktitle = {Proceedings of the 2018 International Conference on Technical Debt},
pages = {57-58},
numpages = {2},
location = {Gothenburg, Sweden},
series = {TechDebt '18}
}

@INPROCEEDINGS{8811932,
  author={LeClair, Alexander and Jiang, Siyuan and McMillan, Collin},
  booktitle={2019 IEEE/ACM 41st International Conference on Software Engineering (ICSE)}, 
  title={A Neural Model for Generating Natural Language Summaries of Program Subroutines}, 
  year={2019},
  volume={},
  number={},
  pages={795-806},
  doi={10.1109/ICSE.2019.00087}
}

@inproceedings{wang2023selfconsistency,
  title={Self-Consistency Improves Chain of Thought Reasoning in Language Models},
  author={Wang, Xuezhi and Wei, Jason and Schuurmans, Dale and Le, Quoc and Chi, Ed H and Narang, Sharan and Chowdhery, Aakanksha and Zhou, Denny},
  booktitle={International Conference on Learning Representations (ICLR)},
  year={2023},
  url={https://arxiv.org/abs/2203.11171}
}

@misc{chen2023usc,
      title={Universal Self-Consistency for Large Language Model Generation}, 
      author={Xinyun Chen and Renat Aksitov and Uri Alon and Jie Ren and Kefan Xiao and Pengcheng Yin and Sushant Prakash and Charles Sutton and Xuezhi Wang and Denny Zhou},
      year={2023},
      eprint={2311.17311},
      archivePrefix={arXiv},
      primaryClass={cs.CL},
      url={https://arxiv.org/abs/2311.17311}, 
}

@inproceedings{taubenfeld2025cisc,
   title={Confidence Improves Self-Consistency in LLMs},
   url={http://dx.doi.org/10.18653/v1/2025.findings-acl.1030},
   DOI={10.18653/v1/2025.findings-acl.1030},
   booktitle={Findings of the Association for Computational Linguistics: ACL 2025},
   publisher={Association for Computational Linguistics},
   author={Taubenfeld, Amir and Sheffer, Tom and Ofek, Eran and Feder, Amir and Goldstein, Ariel and Gekhman, Zorik and Yona, Gal},
   year={2025},
   pages={20090-20111} }

@inproceedings{prasad2024scpo,
title={Self-Consistency Preference Optimization},
author={Archiki Prasad and Weizhe Yuan and Richard Yuanzhe Pang and Jing Xu and Maryam Fazel-Zarandi and Mohit Bansal and Sainbayar Sukhbaatar and Jason E Weston and Jane Yu},
booktitle={Forty-second International Conference on Machine Learning},
year={2025},
url={https://openreview.net/forum?id=94G4eL3RWi}
}

@misc{zhou2025scir,
      title={Self-Consistency of the Internal Reward Models Improves Self-Rewarding Language Models}, 
      author={Xin Zhou and Yiwen Guo and Ruotian Ma and Tao Gui and Qi Zhang and Xuanjing Huang},
      year={2025},
      eprint={2502.08922},
      archivePrefix={arXiv},
      primaryClass={cs.AI},
      url={https://arxiv.org/abs/2502.08922}, 
}

@article{DBLP:journals/corr/abs-2107-07112,
  author       = {Ensheng Shi and
                  Yanlin Wang and
                  Lun Du and
                  Junjie Chen and
                  Shi Han and
                  Hongyu Zhang and
                  Dongmei Zhang and
                  Hongbin Sun},
  title        = {Neural Code Summarization: How Far Are We?},
  journal      = {CoRR},
  volume       = {abs/2107.07112},
  year         = {2021}
}

@inproceedings{DBLP:conf/icse/HaiducAM10,
  author       = {Sonia Haiduc and
                  Jairo Aponte and
                  Andrian Marcus},
  title        = {Supporting program comprehension with source code summarization},
  booktitle    = {{ICSE} {(2)}},
  pages        = {223--226},
  publisher    = {{ACM}},
  year         = {2010}
}

@article{DBLP:journals/corr/abs-2006-03654,
  author       = {Pengcheng He and
                  Xiaodong Liu and
                  Jianfeng Gao and
                  Weizhu Chen},
  title        = {DeBERTa: Decoding-enhanced {BERT} with Disentangled Attention},
  journal      = {CoRR},
  volume       = {abs/2006.03654},
  year         = {2020}
}

@article{DBLP:journals/corr/abs-2409-14368,
  author       = {Aysa Xuemo Fan and
                  Arun Balajiee Lekshmi Narayanan and
                  Mohammad Hassany and
                  Jiaze Ke},
  title        = {Evaluating the Quality of Code Comments Generated by Large Language
                  Models for Novice Programmers},
  journal      = {CoRR},
  volume       = {abs/2409.14368},
  year         = {2024}
}

@article{DBLP:journals/tosem/DongJJL24,
  author       = {Yihong Dong and
                  Xue Jiang and
                  Zhi Jin and
                  Ge Li},
  title        = {Self-Collaboration Code Generation via ChatGPT},
  journal      = {{ACM} Trans. Softw. Eng. Methodol.},
  volume       = {33},
  number       = {7},
  pages        = {189:1--189:38},
  year         = {2024}
}

@misc{code-understanding,
  author       = {Anonymous},
  title        = {Code Understanding Repository},
  howpublished = {\url{https://anonymous.4open.science/r/Code-Understanding-4F19/}},
  note         = {Accessed: May 2026}
}

@inproceedings{DBLP:conf/kbse/LopezWCS22,
  author       = {Jos{\'{e}} Antonio Hern{\'{a}}ndez L{\'{o}}pez and
                  Martin Weyssow and
                  Jes{\'{u}}s S{\'{a}}nchez Cuadrado and
                  Houari A. Sahraoui},
  title        = {AST-Probe: Recovering abstract syntax trees from hidden representations
                  of pre-trained language models},
  booktitle    = {{ASE}},
  pages        = {11:1--11:11},
  publisher    = {{ACM}},
  year         = {2022}
}

@inproceedings{DBLP:conf/blackboxnlp/TroshinC22,
  author       = {Sergey Troshin and
                  Nadezhda Chirkova},
  title        = {Probing Pretrained Models of Source Codes},
  booktitle    = {BlackboxNLP@EMNLP},
  pages        = {371--383},
  publisher    = {Association for Computational Linguistics},
  year         = {2022}
}

@inproceedings{DBLP:conf/icse/NorthAB25,
  author       = {Marc North and
                  Amir Atapour{-}Abarghouei and
                  Nelly Bencomo},
  title        = {Beyond Syntax: How Do LLMs Understand Code?},
  booktitle    = {NIER@ICSE},
  pages        = {86--90},
  publisher    = {{IEEE}},
  year         = {2025}
}

@inproceedings{peitek2022correlates,
author = {Peitek, Norman and Bergum, Annabelle and Rekrut, Maurice and Mucke, Jonas and Nadig, Matthias and Parnin, Chris and Siegmund, Janet and Apel, Sven},
title = {Correlates of programmer efficacy and their link to experience: a combined EEG and eye-tracking study},
year = {2022},
isbn = {9781450394130},
publisher = {Association for Computing Machinery},
address = {New York, NY, USA},
url = {https://doi.org/10.1145/3540250.3549084},
doi = {10.1145/3540250.3549084},
booktitle = {Proceedings of the 30th ACM Joint European Software Engineering Conference and Symposium on the Foundations of Software Engineering},
pages = {120-131},
numpages = {12},
location = {Singapore, Singapore},
series = {ESEC/FSE 2022}
}

@article{zhang2024eyetrans,
author = {Zhang, Yifan and Li, Jiliang and Karas, Zachary and Bansal, Aakash and Li, Toby Jia-Jun and McMillan, Collin and Leach, Kevin and Huang, Yu},
title = {EyeTrans: Merging Human and Machine Attention for Neural Code Summarization},
year = {2024},
issue_date = {July 2024},
publisher = {Association for Computing Machinery},
address = {New York, NY, USA},
volume = {1},
number = {FSE},
url = {https://doi.org/10.1145/3643732},
doi = {10.1145/3643732},
journal = {Proc. ACM Softw. Eng.},
month = jul,
articleno = {6},
numpages = {22}
}

@article{wallace2025programmer,
author = {Wallace, Robert and Bansal, Aakash and Karas, Zachary and Tang, Ningzhi and Huang, Yu and Jia-Jun Li, Toby and McMillan, Collin},
title = {Programmer Visual Attention During Context-Aware Code Summarization},
year = {2025},
issue_date = {May 2025},
publisher = {IEEE Press},
volume = {51},
number = {5},
issn = {0098-5589},
url = {https://doi.org/10.1109/TSE.2025.3554990},
doi = {10.1109/TSE.2025.3554990},
journal = {IEEE Trans. Softw. Eng.},
month = may,
pages = {1524-1537},
numpages = {14}
}

@misc{wan2025reasoningawareselfconsistencyleveraging,
      title={Reasoning Aware Self-Consistency: Leveraging Reasoning Paths for Efficient LLM Sampling}, 
      author={Guangya Wan and Yuqi Wu and Jie Chen and Sheng Li},
      year={2025},
      eprint={2408.17017},
      archivePrefix={arXiv},
      primaryClass={cs.CL},
      url={https://arxiv.org/abs/2408.17017}, 
}
\end{document}